\documentclass[12pt,aps,prb,floats,eqsecnum,amsmath,graphicx,euscript,amsmath,amssymb,onecolumn]{revtex4}
\usepackage[dvips]{graphicx}
\newlength{\intwidth}
\DeclareRobustCommand{\fpint}[2]
  {\mathop{%
     \text{%
       \settowidth{\intwidth}{$\int$}%
       \makebox[0pt][l]{\makebox[\intwidth]{\mbox{\footnotesize\boldmath ${-}$}}}%
       $\int_{#1}^{#2}$}}}
\renewcommand{\min}{\mathop{\rm min}\nolimits}

\def\lapprox{\,\raise0.4ex\hbox{$<$}\kern-0.8em\lower0.7ex\hbox{$\sim$}\,}
\def\gapprox{\,\raise0.4ex\hbox{$>$}\kern-0.8em\lower0.7ex\hbox{$\sim$}\,}
\def\lg{\,\raise0.5ex\hbox{\footnotesize $<$}\kern-0.8em\lower0.5ex\hbox{\footnotesize  $>$}\,}
\def\gl{\,\raise0.5ex\hbox{\footnotesize $>$}\kern-0.8em\lower0.5ex\hbox{\footnotesize  $<$}\,}

\begin{document}
\title{Collective excitations in a magnetically doped quantized Hall ferromagnet}
\author{S. Dickmann$^{1,2}$, V. Fleurov$^{1,3}$ and K. Kikoin$^{1,3,4}$}
\affiliation{ $^1$Max-Planck-Institut f\"ur Physik Komplexer Systeme,
N\"othnitzer Str. 38, D-01187 Dresden, Germany.\\
    $^2$Institute for Solid
State Physics of RAS, Chernogolovka 142432, Moscow District, Russia.\\
    $^3$School of Physics and Astronomy, Beverly and Raymond Sackler Faculty of
Exact Sciences, Tel Aviv University, Tel Aviv 69978,
Israel. \\
    $^4$Department of Physics, Ben-Gurion University of the Negev,\\
Beer-Sheva 84105, Israel.}
\begin{abstract}
A theory of collective states in a magnetically quantized
two-dimensional electron gas (2DEG) with half-filled Landau level
(quantized Hall ferromagnet) in the presence of magnetic $3d$
impurities is developed. The spectrum of bound and delocalized
spin-excitons as well as the renormalization of Zeeman splitting of
the impurity $3d$ levels due to the indirect exchange interaction
with the 2DEG are studied for the specific case of $n$-type GaAs
doped with Mn where the Land\'e $g$-factors of impurity and 2DEG
have opposite signs. If the sign of the 2DEG $g$-factor is changed
due to external influences, then impurity related transitions to new
ground state phases, presenting various spin-flip and skyrmion-like
textures, are possible. Conditions for existence of these phases are
discussed.

\end{abstract}
\maketitle
\newpage
\section{Introduction}

In a strong magnetic field the two-dimensional electron states in
semiconductor heterostructures$\,$\cite{an82} transform into Landau
states with a completely discrete energy spectrum. This
diamagnetically quantized two-dimensional electron gas (2DEG)
possesses many remarkable features including Quantum Hall
effect.\cite{QHE}  The role of impurities in the thermodynamic,
optical and transport properties of 2DEG is extremely important.
Among many facets of this problem we choose for discussion in this
paper the formation of impurity related collective excitations in a
magnetically doped quantized 2DEG in the case of odd integer filling
factor $\nu=2n\!+\!1$. In a pristine state 2DEG with odd $\nu$ is in
a Quantuzed Hall Ferromagnet (QHF) regime with nondegenerate ground
state characterized by the total spin quantum number $S = N_\phi/2$
and maximum spin projection $S_z = S$. ($N_\phi$ is the
magnetic-flux-quanta number.) Different branches of the excitons are
well distinguishable among the low-energy excitations. They are
classified as spin waves (spin excitons), magnetoplasmons or
multi-exciton states depending on the spin and orbital quantum
numbers.\cite{Lelo80,Bychok81,KH84,dz83-84,pi92,gi05,di05,va06}
Besides, low-lying collective half-integer-spin fermionic states
(trions, skyrmions,...) may be formed in a QHF under certain
circumstances.\cite{so93,fe94,ba95,ma96,pa96,ip97,ku99,di02}
Magnetic impurities are characterized by their own spectrum of {\em
local} spin excitations, and one can anticipate a strong interplay
between local and collective excitations in a magnetically doped
QHF.

It is known that the influence of impurities on the discrete
spectrum of quantized Landau electrons in a 2DEG has many specific
features. Even such a basic property, as the interaction of a 2DEG
with neutral short range impurities is far from being
trivial.\cite{Ando74,BME78,aag93} Only the Landau states with a
finite probability density on the scatterer locations interact with
impurities. This means that the whole set of Landau states breaks
down into two groups: the major part of the Landau levels (LLs) is
not affected by the impurity scattering, and the states having a
nonzero scattering amplitude on an impurity form a separate system
of bound Landau states in the energy gaps between the free LLs.

To be more specific, we consider a 2DEG formed in the $n$-type
GaAs/GaAlAs heterostructures doped with transition metal (TM)
impurities. The reason for this choice is that the technology of
(Ga,Mn)As epilayers is well developed, and  the QHF regime is
achieved experimentally in GaAs based heterostrures. As a rule,
transition metal ions substitute for the metallic component of the
binary II-VI and III-V semiconductors.\cite{LB89,KF94,Zung86a}
The influence of isolated TM impurities on the spectrum of the
Landau states was investigated in Ref. \onlinecite{dfvk02}. It was
shown that the resonance scattering in the $d$-channel is in many
respects similar to that of the short range impurity scattering in
the $s$-channel.\cite{Ando74,BME78,aag93} The symmetry selection
rules for the resonance $d$-waves in the cylindrical (symmetric)
gauge pick up the Landau states with the orbital number $m=0$ (in
the symmetric gauge). These states are the same states that are
involved in the $s$-scattering by the impurities with a short range
scattering potential.\cite{aag93} Besides, this scattering is spin selective
in magnetically quantized 2DEG.

It should be emphasized that in the problem under consideration
the criterion of
isolated impurities acquires a specific feature. In fact the Mn
concentration range, where our theory is applicable, is limited from
below by technological capabilities and from above by the obvious
requirement that the impurity induced disorder {\em does not destroy
collective excitonic states}. So, the relevant interval of bulk Mn
concentrations is $10^{13}\,$cm$^{-3} < n_{\rm Mn}\lapprox
10^{15}\,$cm$^{-3}$. Here the upper limit corresponds to the 2D
concentration of $10^9\,$cm$^{-2}$ which in our case is actually
well below the Landau band capacity $N_\phi$ at $B\sim 10$T that
equals to the the 2D electron number on the upper (half-filled) LL.
One may expect that the above mentioned classification of excitonic
states is valid only at the bulk concentration $n_{\rm Mn}\lapprox
10^{15}$ cm$^{-3}$, which is much less than in the materials used
for creation of dilute magnetic semiconductors.\cite{McD}

We calculate in this paper spectra of bound and continuous
collective excitations related to magnetic impurities. When studying
the influence of magnetic impurities on the excitonic spectrum of
2DEG, a distinction between the negative and positive signs of the
gyromagnetic ratio of 2DEG electrons $g_{\mbox{\tiny 2DEG}}$ should
be also mentioned. It will be shown that in the conventional
situation of negative $g_{\mbox{\tiny 2DEG}}$ the interaction with
magnetic impurity lowers the ground state energy due to effectively
antiferromagnetic character of the effective indirect exchange. This
results in formation of a set of bound and delocalized collective
excitations presenting combined modes classified by a change in the
total spin number $S_z$. When $g_{\mbox{\tiny 2DEG}}>0$, so that the
$g$ factors of both subsystems (2DEG electrons and impurities) have
the same sign, magnetic impurities may form bound states in the gap
below the spin exciton continuum and even initiate a global
reconstruction of the QHF ground state.

\section{Model Hamiltonian}

Following Ref. \onlinecite{KF94}, we describe the electron
scattering on a TM impurity in semiconductor within the framework of
the Anderson impurity model Hamiltonian$\,$\cite{Anders61}
generalized for the case of multicharged impurity states in
semiconductors.\cite{HA76,FK76,FK3} According to this model, the
principal source of magnetic interaction is the resonance scattering
of conduction electrons on the $d$-electron levels of TM impurity in
the presence of a strong on-site Coulomb interaction $U$. Due to
this interaction, the local moment of TM impurity survives in the
crystalline environment, and `kinematic' indirect exchange
interaction between the conduction and impurity electrons arises in
the second order in the $s$-$d$-hybridization parameter, even in the
absence of a direct exchange.

The generic Hamiltonian describing the QHF regime in a magnetically
doped semiconductor is
\begin{equation}\label{1.00}
\hat H = \hat H_d+\hat H_s+\hat H_t+\hat H_{sd}.
\end{equation}
Here $\hat H_d = \sum_i \hat H_{di}$ describes the TM impurities on
the sites $i$, $\hat H_s$ is related to the  band electrons on the
LLs, and $\hat H_t$ is responsible for hybridization between the
impurity $d$-electrons and Landau electrons. Eventually, it is this
hybridization that generates coupling between collective modes in
2DEG and localized spin excitations on the impurity sites. In our
extremely weak doping regime both the direct and indirect
interactions between magnetic impurities are negligible. Each
magnetic scatterer may be considered independently, and it is
convenient to choose the symmetric gauge ${\bf A} = (-\frac{B}{2}y,
\frac{B}{2}x,0)$ with the quantum numbers $\lambda = (n,m)$ for the
Landau electrons hybridized with the atomic $d$-electrons centered
around the site $i$ positioned in the center of coordinates. The
Coulomb interaction is taken into account in the impurity and in the
band electron subsystems. Besides the direct Coulomb interaction
between the $d$ and $s$ electrons described by the last term in the
Hamiltonian (\ref{1.00}) is added to the conventional impurity
Hamiltonian (cf. Ref. \onlinecite{Anders61}) described by the first
and third terms. All additional interactions will be discussed below
in detail.

Substitutional Mn impurity in GaAs retains all its five $d$ electrons
due to a special stability of the half filled $3d$ shell. In the
$p$-type GaAs the electrically neutral state of Mn in Ga position is
Mn$^{(3+)}(3d^5+$hole), where the hole is bound on the relatively
shallow acceptor level near the top of the valence band, whereas the
occupied $d$-electron levels are deep in the valence
band.\cite{KF94,Zung86a,KIP,Zung04} In the $n$-type heterostructures
these acceptor states are overcompensated, and the chemical potential
is pinned on one of the lowest Landau levels in the conduction band.
Since we are interested only in the low-energy excitations above the
ground state of $n$-type system, Mn impurities will be considered as
the Mn$^{(3+)}(3d^5)$ ions in the subsequent calculations.

\subsection{Single-orbital model. Spin-selective hybridization.}
\label{II.A}

One may significantly simplify the calculation of the spectrum of
excitations by reducing the general Hamiltonian (\ref{1.00}) to the
form, in which only the terms relevant to the calculation of desired
collective states are present. As a result of this simplification
outlined below in Subsection \ref{II.B} one arrives at the {\em
single-orbital, single Landau band hybridization Hamiltonian}, which
explicitly takes into account the Hund rule governing the high-spin
states $3d^5$ and $3d^6$ of the Mn $3d$ shell (the state $3d^4$ is
proved to be irrelevant in our specific case of Mn in GaAs lattice, see
below Fig. 1). These impurity states are characterized by the maximum
total spin quantum numbers $S^{(d)} = 5/2$ at $3d^5$, and $S^{(d)} = 2$
at $3d^6$, and the effective Hamiltonian $\hat H$ is defined in the
charge sector $\left\{|d^5,N\rangle, |d^6, N - 1\rangle\right\}$ of
states with variable number $N$ or $N - 1$ of the electrons on the
highest $n$-th LL (of course in our case $N \approx N_\phi$). The
Hamiltonian now reads
\begin{equation}\label{H}
\hat H=\sum_{\sigma} \epsilon_{d\sigma} \hat n_{\gamma_0\sigma} + U\hat
n_{\gamma_0\uparrow}\hat n_{\gamma_0\downarrow} + \sum_{m\sigma}
\varepsilon_{n\sigma} a^\dagger_{nm\sigma} a_{nm\sigma} + \hat H'_{\rm
Coul} + \hat H_t\,.
\end{equation}
where the impurity Hamiltonian $\hat H_d$ of Eq. (\ref{1.00}) is
represented in by the two first terms, in which $\hat
n_{\gamma_0\sigma} = c^\dag_{\gamma_0\sigma} c_{\gamma_0\sigma}$, and
$c^{\dag}_{\gamma_0\sigma}$ is the creation operator for the
$d$-electron at the orbital $\gamma_0$ with the spin $z$-component
$\sigma$. The notation $\gamma_0$ designates the only $3d$-orbital with
the $Y_{02} \sim 3 z^2 - r^2$ symmetry, which effectively couples with
the $m=0$ state of the LL.\cite{dfvk02} The parameter $U$ characterizes
Coulomb and exchange interactions determining the addition energy for
the transition $3d^5 \to 3d^6$. The third term in Eq. (\ref{H}) is the
Hamiltonian of noninteracting Landau electrons where
$a^\dag_{nm\sigma}$ is the creation operator for the $(n,m,\sigma)$
Landau state.  The most of interaction components are included in $\hat
H'_{\rm Coul}$. This term does not include only the $d$-$d$ interaction
parametrized by $U$ and the last term ${\hat H}_t$. The latter
generically is also the part of Coulomb interaction between impurity
and Landau electrons which intermixes impurity and Landau orbitals .
However, in our case ${\hat H}_t$ acquires the form of single-electron
hybridization operator [see discussion after Eq. (\ref{H_sd})],
\begin{equation}\label{H_t}
\hat H_t = \sum_{\sigma} W_{n0} a^\dagger_{n0\sigma} c_{\gamma_0\sigma}
+ H.c.
\end{equation}
 As was mentioned above, this operator is responsible
for the resonance orbital-selective scatterings in QHF. It includes
hybridization of the impurity electron with the 2DEG electrons within the
$n$-th LL. This means that only the
influence of impurity on the intra-LL excitations (of the
spin-wave type) is taken into account. The hybridization with the
states with $n'\!\neq\! n$ describing the processes with energy change
$\hbar\omega_c$ or higher is omitted.

In the absence of interaction term $\hat H'_{\rm Coul}$, the
Hamiltonian (\ref{H}) acts in the subspace
\begin{equation}\label{i-states}
|d^5,s;\mbox{vac}\rangle, |d^6,s + \frac{1}{2};
a_{n0\uparrow}|\mbox{vac}\rangle\ \mbox{and}\ |d^5,s + 1;
a^\dag_{n0\downarrow} a_{n0\uparrow}|\mbox{vac}\rangle,
\end{equation}
where the fully polarized 2DEG without impurity is chosen to be the
`vacuum' state $|\mbox{vac}\rangle = \uparrow,\uparrow,
...\uparrow\rangle$. Therefore $a_{nm\uparrow}^\dag|\mbox{vac}
\rangle = a_{nm\downarrow} |\mbox{vac}\rangle\equiv 0$. We represent
the total spin component as $S_z = \frac{N_{\phi}}{2} + s$. Thus we
characterize the states in the set (\ref{i-states}) by the quantum
number $S_z$. It is important that only $S_z = S_z^{(s)} +
S_z^{(d)}$ is an exact spin quantum number in our system.
Separately, the Hamiltonian (\ref{H}) commutes neither with the spin
component $S_z^{(s)}$ of the LL electrons nor with the impurity spin
component $S_z^{(d)}$. Equally, it does not commute with the total
spin ${\bf S}^2$ and with the spins $({\bf S}^{(s)})^2$ and $({\bf
S}^{(d)})^2$ (see Appendix \ref{B}). The number $s$ in the set
(\ref{i-states}) changes within the interval $- \frac{5}{2} \le s <
\frac{5}{2}$. It is convenient to choose the state $|d^5,
\frac{5}{2}; \mbox{vac}\rangle$ as a reference point (`global
vacuum'). This state is not mixed with any other state of the system
by the operator (\ref{H_t}) so that it enters the set of eigenstates
of the Hamiltonian (\ref{H}), although at $g_{\rm 2DEG} < 0$ it is
one of the excited states of the system.

Within a given `triad' (\ref{i-states}), i.e. at a given $s$, the
operator (\ref{H_t}) intermixes these basis states. The
corresponding non-diagonal matrix elements are
$\langle\mbox{vac};s,d^5|\hat H_t|d^6,s +
\frac{1}{2};a_{n0\uparrow}|\mbox{vac}\rangle$ and
$\langle\mbox{vac}|a_{n0\uparrow}^\dag;s + \frac{1}{2},d^6|\hat
H_t|d^5,s + 1; a^\dag_{n0\downarrow} a_{n0\uparrow}|
\mbox{vac}\rangle$, where the bra- and ket-vectors are appropriately
normalized. Therefore, {\em for any given quantum number} $S_z =
\frac{N_\phi}{2} + s$ the magnetic impurity scattering problem can
be effectively described in terms of a {\em single-orbital} impurity
model that involves only one or two $d\gamma_0$-electrons. The
single-orbital basis
\begin{equation}\label{3basis}
|s_-;\mbox{vac}\rangle,\quad
|s_0;a_{n0\uparrow}|\mbox{vac}\rangle\quad\mbox{and} \quad
|s_+;a^\dag_{n0\downarrow}a_{n0\uparrow}|\mbox{vac}\rangle
\end{equation}
arises instead of the original multi-electron basis (\ref{i-states})
where the indices $(-,\,0,\,+)$ label the bare energies $E_{s_-}$,
$E_{s_0} = E_{s_-} + U + \epsilon_{d\uparrow} -
\varepsilon_{n\uparrow}$ and $E_{s_+} = E_{s_-} + (g_i - g_{\mbox{\tiny
2DEG}})\mu_BB$. Here $g_i\mu_BB = \epsilon_{d\uparrow} -
\epsilon_{d\downarrow}$, and $g_{\mbox{\tiny 2DEG}}\mu_BB =
\varepsilon_{n\uparrow} - \varepsilon_{n\downarrow}$ are Zeeman
energies for impurity and 2DEG, respectively. The two states
$E_{s_\pm}$ form a Zeeman doublet for a given $s$, while the state
$E_{s0}$ becomes resonant with the LL continuum.

As a result of this mapping, where only one component $\gamma_0$ of
the normalized multi-electron states is responsible for the
hybridization, the ${\hat H}_t$ operator can be redefined for each
triad (\ref{3basis}) as \vspace{-2mm}
\begin{equation}\label{H_t-mod}
\hat{\cal H}_{t}(S_z) = V\left[\beta_\uparrow(s) c_{\uparrow}^\dag
a_0 + \beta_\downarrow(s) c_{\downarrow}^\dag b_0\right] +
\;\;\mbox{H.c.}
\end{equation}
It becomes thereby {\em spin-selective}. The shorthand notation $a_0
\equiv a_{n0\uparrow}$, $b_0 \equiv a_{n0\downarrow}$ is used here and
below; $c^{\dag}_{\uparrow/\downarrow}$ is the creation Fermi operator
for the $s_{\pm}$ impurity states $|s_\pm;\mbox{vac}\rangle =
c_{\uparrow/\downarrow}^\dag|\mbox{vac}\rangle$ and
$|s_0;\mbox{vac}\rangle = c_{\uparrow}^\dag c_{\downarrow}^\dag
|\mbox{vac}\rangle$. The Clebsch-Gordan coefficients $\beta_\sigma(s)$
reflect normalization of eigenvectors (\ref{i-states}) by replacing
them with normalized single-orbital basis (\ref{3basis}). For
$s=-5/2,-3/2,-1/2,1/2,3/2$ we have
\begin{equation}\label{beta}
\beta_\uparrow=\sqrt{\frac{1}{2}-\frac{s}{5}},\quad
\beta_\downarrow=\sqrt{\frac{7}{10}+\frac{s}{5}}\,.
\end{equation}
The highest state in the bare Zeeman ladder
$|\left(\frac{5}{2}\right)_-;\mbox{vac}\rangle \equiv
|d^5,\frac{5}{2};\mbox{vac}\rangle$ remains nonhybridized.

Unlike the original Anderson model$\,$ \cite{Anders61} the mixing
coefficient, $V \equiv W_{n0}$ in our particular case, arises as a
non-diagonal component of the $s$-$d$ Coulomb interaction (see
Subsections \ref{II.C} and \ref{II.D} for further discussion).

\subsection{Description of the employed simplifications}
\label{II.B}

Here we list the simplifications which have allowed us to reduce Eq.
(\ref{1.00}) to the Hamiltonian (\ref{H}) with following change $\hat
H_t\!\to\hat{\cal H}_t$, and to apply to our system.

The first simplification exploits the fact that the characteristic
Coulomb energy of Landau electrons $E_{\rm C}=\alpha e^2/\kappa l_B$ is
considered to be small in the QHF regime as compared to the cyclotron
energy $\hbar\omega_c$. Here $\alpha$ is the average form-factor
related to the finite thickness of the 2DEG
($0.3\lapprox\alpha\!<\!1$). In the $E_{\rm C}\!\ll \hbar\omega_c$
limit  one may neglect the LL mixing. Besides, it implies that in our
case the energies of collective excitations are smaller than
$\hbar\omega_c$. \vspace{3mm}
\begin{figure}[h]
\includegraphics*[width=.7\textwidth,angle=0]{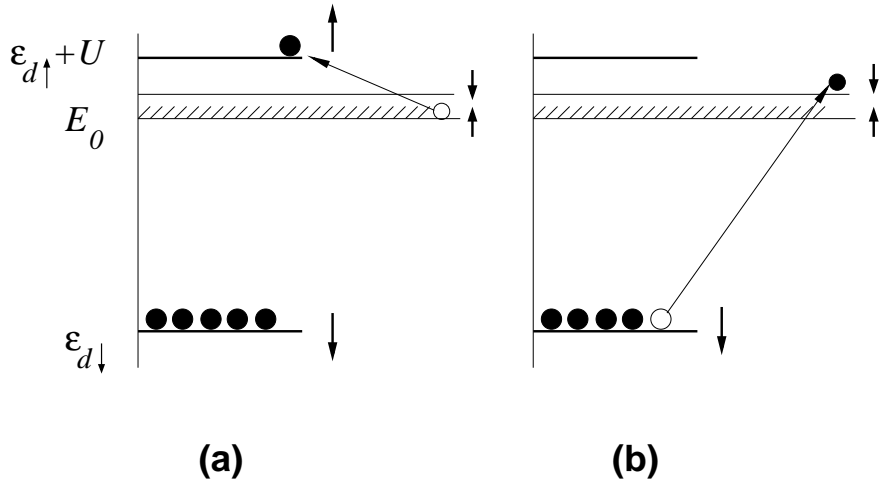}\vspace{-5mm}
\caption{Allen reactions that involve an additional electron (a)
or hole (b) in the impurity site. $\epsilon_{d\downarrow}$ and
$\varepsilon_{d\uparrow} + U$ are addition energies for the 5-th
and 6-th electrons in the $3d$-shell of Mn ion in accordance with
Eqs. (2.8). The ground state with the energy $E_0$ corresponds to
the completely occupied lowest Landau subband. Spins of Mn $3d$
shell and occupied Landau subband are antiparallel because of the
different signs of $g$ factors for Mn and 2DEG in GaAs.}
\label{f.1}
\end{figure}

The second simplification is related to the `deepness' of the
$3d$-levels of a neutral Mn impurity relatively to the bottom of
conduction band in GaAs. We know from the previous
studies$\,$\cite{dfvk02} that the scattering potential created by a TM
impurity for the Landau electrons is generated by the $s$-$d$
hybridization. It has the resonance character, and the spin selective
scattering becomes strong when one of the impurity $3d$ levels is close
to the LLs of conduction electrons. The process of $s$-$d$
hybridization may be represented by the so called `Allen
reactions'\cite{KF94,Zung86a,Allen} (see Fig. \ref{f.1}) \vspace{-4mm}
$$
\begin{array}{ccr}
{}\qquad{}\qquad{}\qquad{}\qquad{}\qquad{}\qquad{}&3d^5   \to 3d^6 +
h\,,& \qquad{}\qquad{}
\qquad{}\qquad{}\qquad{}\qquad{}\qquad{}\qquad{}\quad{}\mbox{(2.8a)}
\\
{}\qquad{}\qquad{}\qquad{}\qquad{}\qquad{}\qquad{}&3d^5 \to 3d^4 +
e\,.& \qquad{}\qquad{}\qquad{}\qquad{}\qquad{}\qquad{}\qquad{}
\qquad{}\quad{}\:{}\mbox{(2.8b)}
\end{array}
$$
\setcounter{equation}{8}

\noindent The first of these reactions describes hopping of an electron
from the filled Landau subband to the impurity $d$-shell, whereas the
second one means hopping of an electron from the $d$ shell to a state
in the empty Landau subband. It is known from the numerical
calculations$\,$\cite{Zung04} that the addition energy for the 6-th
electron in the Mn $3d$ shell (${\rm e_-^{CFR}}$ state in terms of Ref.
\onlinecite{Zung86a}) is in resonance with the states near the bottom
of GaAs conduction band. It really means that the values $U$ and
$\varepsilon_{d\uparrow}$ well compensate each other in the sum
$\varepsilon_{d\uparrow} + U$. So, one may retain only the processes
(2.8a) in $\hat H_t$ and neglect contributions of the $3d^5 \to 3d^4$
ionization.

The third major reduction of the Hamiltonian is the elimination of the
impurity orbital degrees of freedom due to the selection rules for the
$s$-$d$ hybridization matrix elements.\cite{dfvk02} This orbital
selectivity arises, first, because of symmetry reasons since only
electrons with equal axial $m$-numbers in $d$ and LL states are
hybridized. Second, a precondition of the selectivity is related to the
fact that the magnetic length $l_B$ is much larger than the radius
$r_d$ of $3d$-electron state (in the energy scale this condition takes
the form of inequality $U,\varepsilon_{d\sigma} \gg \hbar\omega_c$).
The hybridization integral determined by the overlap of the $d$- and
Landau wave functions behaves as $\sim (r_d/l_B)^m \ll 1$ for $m \neq
0$. All resonance scattering (hybridization) amplitudes with $m\neq 0$
are thus negligibly small, and only the $s$-scattering term ($m = 0$)
can be retained in $H_t$. This explains why only one of the five
$3d$-orbitals, namely $\gamma_0$, is involved in the resonance
interaction with the 2D Landau electrons.

\subsection{Interaction Hamiltonian in excitonic representation}
\label{II.C}
As it was mentioned above, the states of the system are characterized
by the total spin component $S_z$. For a given $S_z$ we may deduce the
effective Hamiltonian
\begin{equation}\label{H-tot}
\hat {\cal H}(S_z)=\hat H_d + \hat H_1^{(s)} + \hat{\cal H}_{t}(S_z) +
\hat H_{s\!-\!s} + \hat H_{s\!-\!d}
\end{equation}
with the single-orbital impurity term $\hat H_d = \epsilon_{d\uparrow}
\hat n_{\uparrow} + \epsilon_{d\downarrow} \hat n_{\downarrow} + U \hat
n_{\uparrow}\hat n_{\downarrow}\;$ (${\hat n}_\sigma \!=c^\dag_\sigma
c_\sigma\!$),~ and with the hybridization term determined by Eq.
(\ref{H_t-mod}).

The remaining terms in the Hamiltonian (\ref{H-tot}) are defined
within the framework of the single-LL approximation for the Landau
electrons.\cite{Lelo80,Bychok81,KH84,di05,so93,fe94,ip97,di02,va06,
dz83-84,Bychok94,FNT,Dick} Although only the states with $m = 0$ in
the LL are involved in the resonance scattering the complete basis
for the description of collective excitations includes all
$m$-orbitals of the LL, and the corresponding Schr\"odinger field
operators should be taken in the form
\begin{eqnarray}\label{Schr-op}
&&  {\hat \Psi}_\uparrow({\bf R})= c_\uparrow\psi_d({\bf R})
+\zeta_s(z)\sum_ma_m{\varphi}_{m}({\bf  r}), \nonumber \\
&&  {\hat \Psi}_\downarrow({\bf R})= c_\downarrow\psi_{d}({\bf R})
+\zeta_s(z)\sum_mb_m{\varphi}_{m}({\bf r}).
\end{eqnarray}
Here the shorthand notation $a_m = a_{nm\uparrow}$, $b_m =
a_{nm\downarrow}$ is used. ${\bf R} = ({\bf r}, z)$ is the 3D
coordinate with the reference point at the impurity site,
$\zeta_s(z)$ is the size-quantized functions of $s$-electrons in the
layer, $\varphi_m$ is the wave function of the $n$-th LL, where
index $m$ in the symmetric gauge changes within the interval $(-n,-n
+ 1,... N_\phi - n - 1)$.

Using the above definitions and Eqs. (\ref{Schr-op}) in the generic
interaction operator
\begin{equation}\label{bareCoul}
\hat H_{\rm Coul} = \frac{1}{2} \sum_{\sigma_1,\sigma_2 = \uparrow,
\downarrow} \int d^3 R_1 d^3 R_2 {\hat \Psi}_{\sigma_2}^\dag({\bf
R}_2) {\hat \Psi}_{\sigma_1}^\dag({\bf R}_1)W\left({\bf R}_1 - {\bf
R}_2\right) {\hat \Psi}_{\sigma_1}({\bf R}_1) {\hat
\Psi}_{\sigma_2}({\bf R}_2)
\end{equation}
[where $W({\bf R})\approx e^2/\kappa R$ at $R\gg r_d$], one may
rewrite the $s$-$s$ and $s$-$d$ Coulomb interactions in the {\em
excitonic representation} (ER).\cite{dz83-84,di05,di02} This
actually means that after substitution of Eqs. (\ref{Schr-op}) into
formula (\ref{bareCoul}) the latter can be expressed in terms of
combinations of various components of the density-matrix operators.
These are so-called ER operators presented in our case only by the
intra-LL set, i.e. by the spin-exciton operators ${\cal Q}^\dag_{\bf
q}$ where an electron is promoted from one spin-sublevel to another
[see Refs. \onlinecite{dz83-84,di05,di02} and Appendix \ref{A} where
the necessary ER equations are given with the reference to our case]
and by operators ${\cal A}^\dag_{\bf q}$ and ${\cal B}^\dag_{\bf q}$
acting within the sublevels $a$ or $b$ (see ibidem). As a result the
Coulomb terms of Eq. (\ref{H-tot}) can be written only by means of
the intra-sublevel operators ${\cal A}^\dag_{\bf q}$ and ${\cal
B}^\dag_{\bf q}$ [their definitions are given by Eq. (A.4) in
Appendix \ref{A}],
\begin{equation}\label{H_ss}
{\hat H}_{s\!-\!s} = \frac{N_\phi}{2} \sum_{\bf q}W_{ss}({ q})
\left({\cal A}^\dag_{\bf q} {\cal A}_{\bf q} + 2 {\cal A}^\dag_{\bf
q}{\cal B}_{\bf q} + {\cal B}^\dag_{\bf q}{\cal B}_{\bf q}\right) -
\frac{1}{2} \left({\cal A}_0 + {\cal B}_0\right)\sum_{\bf q}
W_{ss}(q)\,,
\end{equation}
\begin{equation}\label{H_sd}
{\hat H}_{s\!-\! d} = ({\hat n}_\uparrow + {\hat n}_\downarrow)
\sum_{\bf q} W_{sd}({ q})({\cal A}_{\bf q} + {\cal B}_{\bf q})\,.
\end{equation}
The Coulomb vertices are presented also in Appendix \ref{A} [Eqs.
(\ref{2.7}) and (\ref{coul})].

We neglect in Eq. (\ref{bareCoul}) the direct exchange $s$-$d$
interaction terms (see the next subsection). The mixing operator
$\hat{\cal H}_{t}(S_z)$ in our model Hamiltonian (\ref{H_t-mod})
includes in fact off-diagonal interaction terms from $\hat H_{\rm
Coul}$. Indeed, Coulomb interaction described by the terms $\sim
c^\dag_\downarrow \hat n_\uparrow b_m$ + H.c. and $\sim
c^\dag_\uparrow{\hat n}_\downarrow a_m$ + H.c. induces transitions
adding or removing one electron to the $d$-center in accordance
with the Allen reaction diagrams (2.8). These terms represent the
$s$-$d$ hybridization formally conditioned by the $d$-center
occupation; however, since in our case the reaction (2.8b) is
forbidden, they actually operate as $\sim c^\dag_\downarrow b_m$
+H.c. and $\sim c^\dag_\uparrow a_m$ + H.c. in Eq.
(\ref{H_t-mod}), respectively. [In terms of the
$d^5\!\leftrightarrow d^6$ transitions the hybridization is taken
just in the form of Eq. (\ref{H_t}).]

The single particle Hamiltonian for LL electrons may be also written in
the ER representation,
\begin{equation}\label{H_1}
{\hat H}_1^{(s)} = N_\phi\left[(\varepsilon_n - \varepsilon_{\rm
Z}/2){\cal A}_0 + (\varepsilon_n + \varepsilon_{\rm Z}/2) {\cal
B}_0\right]\,,
\end{equation}
where $\varepsilon_{\rm Z}=|g_{\mbox{\tiny 2DEG}}|\mu_B B$ and
$\varepsilon_n = (n+1/2)\hbar\omega_c$.

\vspace{-1mm}
\subsection{Numerical estimates of the energy parameters}
\label{II.D}

Before turning to our main task, i.e. to the calculation of excitation
spectra, it is worthwhile to evaluate the characteristic energy
parameters related to this problem. We estimate the parameters of 2DEG
in GaAs for the typical value $B=10\,$T of magnetic field. In this
field $E_{\rm C}\! \sim\! 5\,$meV characterizes the Coulomb interaction
(\ref{coul}). Below in our calculation this value is mostly presented
by the spin-exciton mass, which can be estimated empirically, i.e. the
inverse mass is $1/M_{\rm x}\!\sim\! 2\,{}$meV in energy units. The
LLs' spacing is $\hbar\omega_c\!\approx\! 16\,{}$meV, and the Zeeman
splitting between two Landau subbands is $\varepsilon_{\rm Z}\!
\approx\! 0.25\,{}$meV (because $g_{\mbox{\tiny 2DEG}} = g_{\mbox{\tiny
GaAs}} \approx - 0.44$). The Zeeman splitting for Mn ion is $g_i \mu_BB
\approx\!1.1\,{}$meV (because $g_i=g_{\mbox{\tiny Mn}}\approx 2.0$).
The hybridization constant $V$ and the repulsion $U$ are the other
important parameters characterizing the magnetic impurity. It is rather
difficult to extract them from the available experimental data. We can
only roughly estimate the energy $U$ as a distance between the
Mn-related peaks in the density of states of occupied and empty states
in the spectrum of bulk (Ga,Mn)As calculated with an account of the
electron-electron interaction.\cite{Zung04} Such an estimate gives $U
\!\sim 4-4.5{}\,{}$eV. From the same calculations we estimate the
energy difference
\begin{equation}\label{Delta}
\Delta = \epsilon_{d\uparrow} + U - \varepsilon_n + \varepsilon_{\rm
Z}/2,
\end{equation}
which determines the position of the Mn($d^6)$ electron level above the
bottom of Landau band (see Fig. \ref{f.1}a) as $\Delta\lesssim
0.1\,$eV. In order to estimate the parameter  $V$, one should recollect
that the dominating contribution to hybridization integral is given by
the matrix elements of Coulomb interaction, having the form
$Vc^\dag_\uparrow a_0 c^\dag_\downarrow c_\downarrow$ (see above). This
means that $V \sim U r_d^{3/2}\zeta(z_d)/l_B$. Estimating the radius of
the $\psi_d$ function as $r_d\sim 2\,$\AA, and $\zeta(z_d)\sim
0.15\,\mbox{\AA}{}^{-1/2}$ (for the impurity located in the vicinity of
the quantum well bottom), one gets $V \sim 20\,$meV. This gives
$|V|^2/\Delta\sim 4-8\,$meV for the relevant kinematic exchange
parameter. At the same time the direct exchange turns out to be
insignificant. Indeed, one can estimate from Eq. (\ref{bareCoul}) that
the characteristic coupling constants for the terms $\sim
c_\uparrow^\dag c_\downarrow b_{m_1}^\dag a_{m_2}$ and $c_\uparrow^\dag
c_\downarrow^\dag b_{m_1}a_{m_2}$ are of the order of $U r_d^3|
\zeta(z_d)|^2/l_B^2$ being therefore by the factor $\sim \Delta/U$
smaller than $|V|^2/\Delta$.

\section{Collective spin-flip states.
negative $\displaystyle{\mbox{\large\boldmath
$g$}}$$_{\mbox{\scriptsize 2DEG}}$-factor}

The Coulomb interactions ${\hat H}_{s - s}$ and ${\hat H}_{s - d}$
admix the LL states with $m \neq 0$ to the three-state basis
(\ref{3basis}). Instead of triads (\ref{3basis}), the basis
\begin{equation}\label{x-basis}
|s_-;\mbox{vac}\rangle,\quad |s_0;a_m|\mbox{vac}\rangle\quad \mbox{and}
\quad |s_+;{\cal Q}_{\bf q}^\dag|\mbox{vac} \rangle
\end{equation}
contains spin-exciton continua ${\cal Q}_{\bf q}^\dag|\mbox{vac}
\rangle$ attached to the spin-flipped impurity state $s_+$. [The
definition of the spin-exciton creation operator is given by Eq. (A1).]

This set is complete only within the single-orbital
approximation.\cite{foot} At a given $s$ it is convenient to take
the energy $E_{0+}{(s)}$ of the state $|0\rangle = |s_+;\mbox{vac}
\rangle \equiv c_\uparrow^\dag|\mbox{vac}\rangle$ as the reference
point because this state is not affected by the hybridization within
the framework of the single-orbital model. This energy is defined as
$E_{0+}(s) = \langle\mbox{vac}|;s + 1,d^5| \hat{H}|d^5,s +
1;|\mbox{vac}\rangle$ where the Hamiltonian $\hat H$ is given by
Eqs. (\ref{H}) and (\ref{H_t}). For a given $S_z = \frac{N_\phi}{2}
+ s$ we have $E_{0+}(s) = E_{\rm vac} - (\frac{5}{2} - s)g_i\mu_BB$
with $E_{\rm vac}$ defined as the energy of the global vacuum state
$|d^5,5/2; |\mbox{vac}\rangle$. One can check with the help of
expressions (\ref{II.3}) and (\ref{II.4}) in Appendix \ref{B} that
the vectors $|s_0;a_m|\mbox{vac}\rangle$ and ${\cal Q}^\dag_{\bf
q}|0\rangle$ at ${\bf q} \not = 0$ correspond to the definite total
spin state with $S = S_z$, whereas $|s_-;|\mbox{vac}\rangle$ and
${\cal Q}^\dag_0 |0\rangle$ are not characterized by any definite
number $S$.\cite{foot2}

\subsection{Secular equation}
\label{III.A}

Following the above discussion the spin-flip operator may be
represented in the form
\begin{equation}\label{24}
{\hat X}^\dag=c^\dag_\downarrow c_\uparrow-\sum_mD_m c^\dag_\downarrow
a^{}_m + \sum_{\bf q}f({\bf q}){\cal Q}^\dag_{\bf q}.
\end{equation}
The normalizability condition $\langle X|X\rangle < \infty$ for the
bound spin-exciton state $|X\rangle= X^\dag|0\rangle$ then reads
$\sum_m|D_m|^2 + \sum_{\bf q}|f({\bf q})|^2 < \infty$ and the sum
$$
N_b = \sum_{\bf q}|f({\bf q})|^2=\frac{N_\phi}{2\pi} \int d{\bf
q}|f({\bf q})|^2 $$
presenting the contribution of continuous spin excitons into the norm
$\langle X|X\rangle$ becomes thereby an essential characteristic of the
spin-flip excitation. For the regular (normalizable) solutions we
expect $f({\bf q}) \sim N_\phi^{-1/2}$. Besides, singular states, for
which the sum $\sum_{\bf q}|f(q)|^2$ diverges also exist. These states
form continuous spectrum of impurity-related spin-excitons.

The coefficients $D_m$ and $f({\bf q})$ are determined from the
equation
\begin{equation}\label{3.3}
[\hat{\cal H},{\hat X}^\dag]|0\rangle = E|X\rangle\,.
\end{equation}
where the energy $E$ is counted from $E_{0+}$(s). Before turning to the
computation we specify the energy levels of the basis states
(\ref{x-basis}) at $V=0$.  The state $|s_- ;\mbox{vac}\rangle$ has the
energy $E_{0-}(s) = E_{0+}(s) - g_i\mu_BB$. The doubly occupied
impurity state $|d_0;a_m|\mbox{vac}\rangle$ appears due to the charge
transfer with creation of a conventional `hole' in the LL. Its energy
is $E_{d,m}(s) = E_{0+} + {\cal E}_{d,m}$ where
\begin{equation}\label{2.15}
{\cal E}_{d,m} = \epsilon_{d\downarrow} + U + \varepsilon_{\rm Z}/2 -
\varepsilon_n + \epsilon_m + {\cal E}_\infty
\end{equation}
[cf. Eq. (\ref{Delta})]. Here ${\cal E}_\infty = (1/N_\phi)\sum_{\bf
q}W_{ss}(q)$ [see Eq. (\ref{I.7}) for definition of ${\cal E}_\infty$].
This term appears due to the global electroneutrality requirement when
calculating the energy of the hole $a_m| \mbox{vac}
\rangle$.\cite{Bychok81,KH84,so93,di05} The term ${\epsilon}_{m} = -
(2/N_\phi) \sum_{\bf q}h_{m + n,m + n}({\bf q})W_{sd}(q)$ is the
Coulomb interaction energy of the hole $a_m|\mbox{vac}\rangle$ with the
doubly occupied $d$-center [see Eq. (\ref{51}) for functions $h_{ik}$].

Substituting operators (\ref{H-tot}) and (\ref{24}) into Eq.
(\ref{3.3}), projecting the result onto the basis vectors
(\ref{x-basis}) and using Eqs. (\ref{H_t-mod}), (\ref{ortho}),
(\ref{H_ss}), (\ref{H_sd}), (\ref{H_1}) and (\ref{I.1})-(\ref{I.6})
we obtain a closed system of equations for the coefficients $D_m$,
$f({\bf q})$. This system defines the eigenvalues of Eq. (\ref{3.3})
for a given $s$. The symmetry of the problem allows us to look for
the solutions in the form $f({\bf q}) = f_m(q) e^{im\phi}$. Below we
limit ourselves to a study of the isotropic case of $m=0$.
(Discussion of the case $m \neq 0$ may be found in Ref.
\onlinecite{FNT}.) As a result we get $D_m = D_0\delta_{m,0}$ and
our system for given $S_z = \frac{N_\phi}{2} + s$ acquires the
simple form
\begin{eqnarray}\label{equations}
E + g_i\mu_BB& = &\beta_\uparrow(s) V^*D_0\,,\nonumber\\
\displaystyle{(E-{\cal E}_{d,0})D_0}&=&\displaystyle{\beta_\uparrow(s)
V-\beta_\downarrow(s) VN_\phi^{-1/2}\sum_{\bf q}
h_{nn}^*({q})f({q})} \,,\nonumber\\
\displaystyle{\left(E - \varepsilon_{\rm Z} - {\cal E}_q\right)f({q})}&
= &\displaystyle{ - N^{-1/2}_\phi h_{nn}({q})\beta_\downarrow(s) V^*D_0
\,.}
\end{eqnarray}
The energy of the free exciton state ${\cal Q}^\dag_{\bf q}|0\rangle$
is $\varepsilon_{\rm Z} + {\cal E}_q$ [see Eq. (\ref{I.7})].

The collective states localized around a magnetic impurity are
described by solutions of Eq. (\ref{equations}) outside the free
spin-wave band (i.e. in the energy interval $E<\varepsilon_{\rm Z}$ or
$ E > \varepsilon_{\rm Z}\! + \!{\cal E}_{\infty}$). The corresponding
eigenfunctions are characterized by the regular envelope function
$f_0(q)$. We arrive then at the secular equation
\begin{equation}\label{secular}
\frac{\beta_\downarrow^2(s)}{N_\phi}\sum_{{\bf q}}\frac{|h_{nn}(q)|^2}
{E - \varepsilon_{\rm Z} - {\cal E}_q} + \frac{\beta_\uparrow^2(s)}{E +
g_i\mu_BB} = \frac{E - {\cal E}_{d,0}}{|V|^2}
\end{equation}
for the energy $E$. The first term in the l.h.s. of Eq.
(\ref{secular}), including the sum of spin-exciton propagators,
presents the self energy, which usually arises in the Schr\"odinger
or Lippmann-Schwinger equation describing the perturbation
introduced by a short-range potential into the continuous spectrum.
The prototype of this term in the theory of magnetic defects is the
self energy for localized spin waves in the Heisenberg ferromagnet
with a single substitution impurity.\cite{Wocal} Specific features
of our model are manifested by the energy dependence of impurity
related processes. First, instead of a constant term (inverse
impurity potential) in the r.h.s. of Eq. (\ref{secular}) we have the
inverse resonance potential $|V|^2/(E - {\cal E}_{d,0})$, which
stems from the hybridization between LLs and the 3d-level of
impurity electron. \cite{FK76,HA76} Second, an additional term
describing impurity spin-flip process in terms of the single-orbital
model arises in the l.h.s. of Eq. (\ref{secular}).

\subsection{Spectrum of the localized states}
\label{III.B}

First we carry out a simple study considering solutions of Eq.
(\ref{secular}) in the absence of an exciton band, i.e. by formally
substituting ${\cal E}_q=0$ into Eq. (\ref{secular}). (This is
instructive in order to classify the bound collective states.) We
obtain then a simple algebraic equation with two roots. When solving
this equation we use the sum rule $\sum_{{\bf q}}|h_{nn}(q)|^2 =
N_\phi$ and neglect the energy dependence  in the r.h.s.  due to the
condition ${\cal E}_{d,0} \approx \Delta \gg E$. Each doublet is
bound to its own reference energy $E_{0+}(s)$ in accordance with the
corresponding spin component $S^{(d)}_z = s + 1$ of the Mn$^{(+2)}$
ion. Due to the kinematic exchange (2-nd order spin-flip processes)
each state in the Zeeman grid (lower root of Eq. \ref{secular})
acquires a partner state (upper root of Eq. \ref{secular}), except
for the highest level with $s = 5/2$, which remains intact, because
the spin flip processes are kinematically forbidden for this state.
The level splitting is illustrated by the scheme in Fig. 2. We see
that the kinematic exchange makes the Zeeman states of impurity ion
non-equidistant and an additional multiplet of excited states arises
as a prototype of the bound spin-excitons.
\begin{figure}[h]\begin{center} \vspace{-3.mm}
\includegraphics*[width=.7\textwidth]{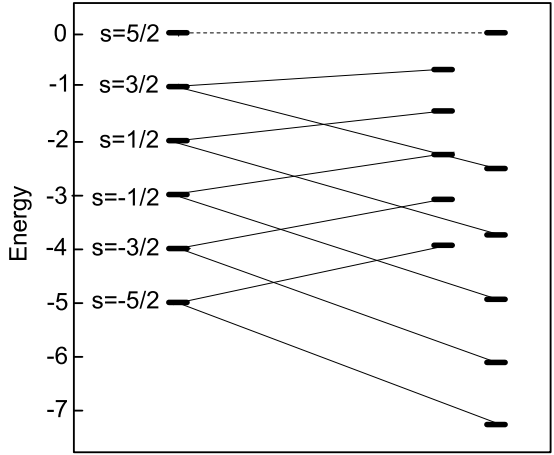}
\end{center}\vspace{-10mm}
\caption{A scheme of the Mn$^{(+2)}$ Zeeman level splitting due to
the kinematic exchange in the absence of exciton dispersion. The
bare Zeeman ladder is shown on the left. Five of six levels in this
grid are shifted down (extreme right column), whereas the $s = 5/2$
level remains not renormalized. Each of these five levels has its
high-energy counterpart. The energy is measured in the $g_i\mu_BB$
units. The following values of the input parameters are chosen:
$\varepsilon_{\rm Z} = 0.2,$ $|V|^2/\Delta = 2$ and ${\cal E}_{d,0}
= \Delta\gg 1$. The factors $\beta_\downarrow^2(s)$ and
$\beta_\uparrow^2(s)$ are presented by Eq. (\ref{beta}).}
\label{f.2}
\end{figure}

Having this classification in mind we turn to calculating the bound
exciton states for a finite dispersion of the free spin waves.
According to the estimates of the model parameters presented in
Subsection \ref{II.C} we solve Eq. (\ref{secular}) for the realistic
conditions $E_{\rm C}\gapprox g_i\mu_BB\gg \varepsilon_{\rm Z}$
whereas the ratio between the energies $E_{\rm C}$ and
$|V|^2/\Delta$ may be arbitrary.

All the generic features of impurity-related states may be seen in
the case of a unit filling where $n=0$ ($\nu = 1$) and we study this
situation in detail. The solutions we are looking for are localized
in the energy interval $|E - {\cal E}_{d,0}|\approx \Delta$ so one
can neglect the energy dependence in the r.h.s. of Eq.
(\ref{secular}). A graphical solution of Eq. (\ref{secular}) is
schematically shown in Fig 3.

Two intersection points labeled as $E_0^{(s)}$ and $E_{\rm x}^{(s)}$
correspond to two discrete solutions. Just as in Fig. 2 this pair of
solutions arises at any $s$ except for $s = 5/2$. The lower solution
with the energy $E_0^{(s)}$ is the state of the Mn$^{(+2)}$ ion with
the spin component $\langle\hat S_z^{(d)}\rangle \approx s$ shifted
downwards from the value $E_{\rm vac} + (s - \frac{5}{2})g_i\mu_BB$
by the effective exchange interaction with the spin-wave continuum
(in this case $D_0 > N_b$). The upper solution corresponds to the
spin-flipped state of the Mn$^{(+2)}$ ion with $\langle {\hat
S}_z^{(d)} \rangle \approx s + 1$ dressed with the spin-wave
localized on the impurity. In this case $\langle \hat S_z^{(s)}
\rangle \approx \frac{N_\phi}{2} - 1$ and $N_b > D_0 $. This bound
exciton state, described semi-phenomenologically in Ref.
\onlinecite{FNT}, is shallow compared with the main characteristic
energy parameter $E_{\rm C}$. Like in many other impurity-related
states in 2DEG\cite{aag93,dfvk02,Demd65} its energy is confined
within the interval $-g_i \mu_B B < E_{\rm x}^{(s)} <
\varepsilon_{\rm Z}$ in the logarithmic vicinity of the bottom of
the delocalized spin-exciton band. Due to this fact one may find the
level position analytically. Using the quadratic approximation for
the exciton dispersion law ${\cal E}_q=q^2/2M_{\rm x}$ and turning
from summation to integration in the l.h.s. of Eq. (\ref{secular})
one has
\begin{figure}[h]\begin{center} \vspace{-8.mm}
\includegraphics*[width=.7\textwidth]{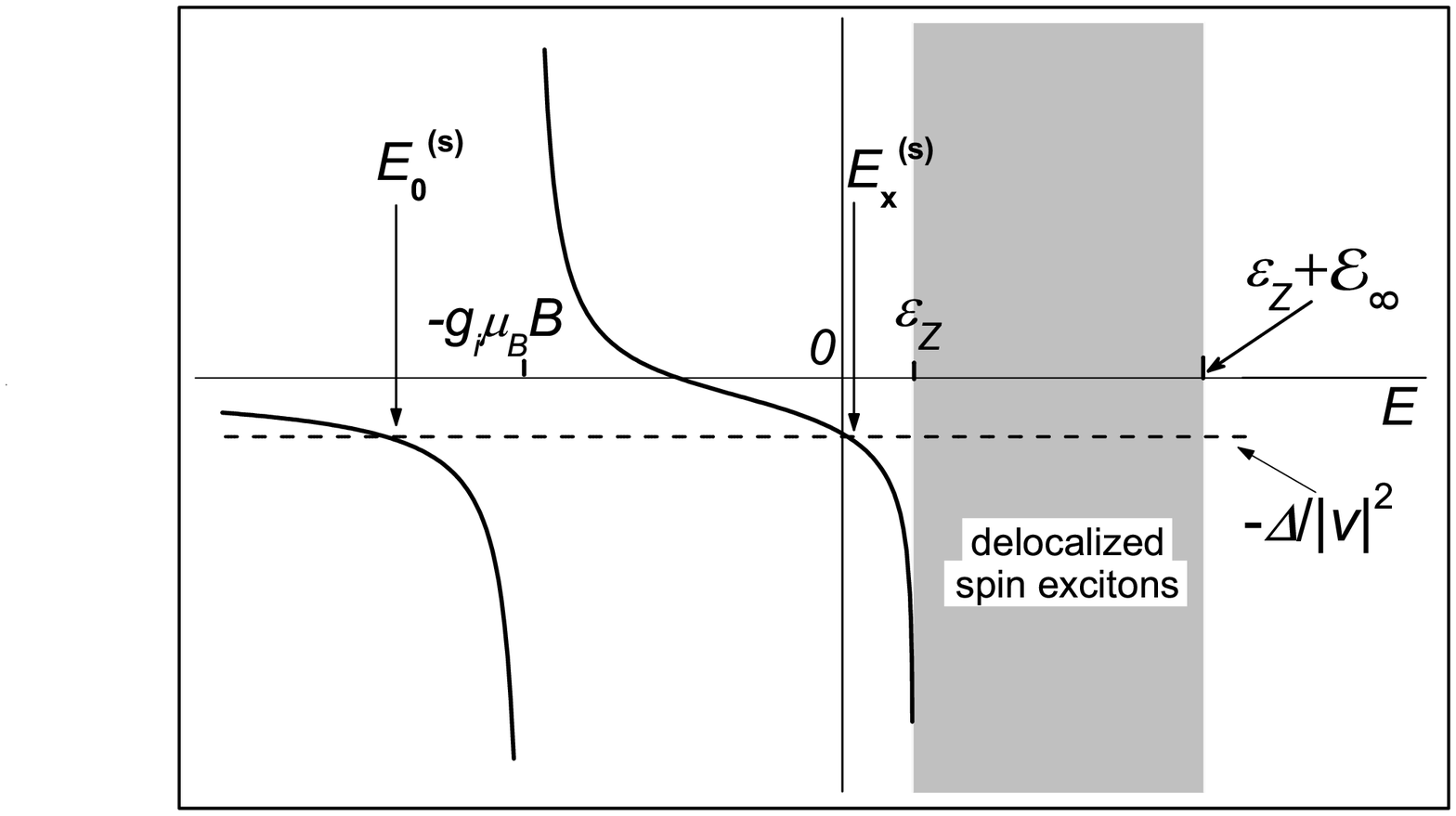}
\end{center}
\vspace{-10mm} \caption{Graphical solution of the secular equation. The
l.h.s. and r.h.s. of Eq. (\ref{secular}) are shown as functions of the
argument $E$ by solid and dashed lines, respectively ($E$-dependence in
the r.h.s. is neglected). Filled area indicates possible values of the
l.h.s. because it belongs to the interval of $E$ where the sum in Eq.
(\ref{secular}) becomes indefinite.} \label{f.3}
\end{figure}
\begin{equation}\label{4.1}
\frac{\beta_\downarrow^2}{N_\phi}\sum_{{\bf
q}\not=0}\frac{|h_{00}(q)|^2} {E - \varepsilon_{\rm Z} - {\cal E}_q}
\approx \beta_\downarrow^2M_{\rm x} \ln\left[\gamma M_{\rm
x}(|\varepsilon_{\rm Z} - E|) \right]
\end{equation}
Here $M_{\rm x}$ is the spin-exciton mass defined as $1/M_{\rm x} =
\int_0^\infty dpp^3{v}_{ss}(p) e^{-p^2/2}/2 \sim E_{\rm C}$ [see Eqs.
(\ref{I.7}) and (\ref{coul}), the $l_B\!=\!1$ units are used.] and
$\gamma = 1.781...$. Then the binding energy
\begin{equation}\label{3.8}
E_{\rm x}^{(s)} = \varepsilon_{\rm Z} - \frac{1}{\gamma M_{\rm
x}}\exp{\left( -\frac{\beta_\uparrow^2}{\beta_\downarrow^2 M_{\rm
x}g_i\mu_BB} - \frac{\Delta}{\beta_\downarrow^2 M_{\rm
x}|V|^2}\right)}.
\end{equation}
is found from Eq. (\ref{4.1}). This result is valid provided at least
one of the two inequalities, $\beta_\downarrow^2M_{\rm x} g_i \mu_BB
\ll \beta_\uparrow^2$ or $\beta_\downarrow^2 M_{\rm x} |V|^2/ \Delta\ll
1$, holds, which is not too strict requirement due to the exponential
smallness of the second term in r.h.s. of Eq. (3.8).

The asymptotic value of the lower state $E_0^{(s)}$ is also easily
found. In the case of strong hybridization $|V|^2/{\Delta}\gg E_{\rm
C}$ one gets $E_0^{(s)} \approx - g_i\mu_BB - {5|V|^2}/{6\Delta}$.
In this asymptotic limit the excitation energy does not depend on
$s$. In the opposite limit $|V|^2/\Delta \ll E_{\rm C}$ we have
$E_0^{(s)} \approx - g_i \mu_BB - {\beta_\uparrow^2(s) |V|^2}/
{\Delta}$.

In the intermediate region $|V|^2/\Delta \sim E_{\rm C}$ Eq.
(\ref{secular}) for $E_0^{(s)}$ can be solved numerically. It is
convenient to rewrite this equation in the dimensionless form
\begin{equation}\label{3.9}
\beta_{\downarrow}^2(s) \int_0^\infty \frac{e^{-q^2/2}q dq} {F^{(s)} -
\xi e(q)} + \frac{\beta_{\uparrow}^2(s)}{F^{(s)} + g} + 1 = 0
\end{equation}
where $\xi = \Delta/M_{\rm x}|V|^2$ is the ratio of the characteristic
Coulomb energy in the Landau band and the characteristic kinematic
exchange energy. The relevant energy parameters in (\ref{secular}) are
redefined as $E = \left(|V|^2/\Delta\right) F^{(s)}(\xi)$, ${}
\varepsilon_{\rm Z} + {\cal E}_q = M^{-1}_{\rm x}e(q)$ and $\,g_i\mu_BB
= \left(|V|^2/\Delta\right)g$. Then the system of localized levels
${\tilde E}_{0,{\rm x}}^{(s)}$ counted from the global vacuum energy is
described by the set of equations
\begin{equation}\label{3.10}
{\tilde E}_{0,{\rm x}}^{(s)} = -g_i\mu_BB(3/2-s) + F^{(s)}_{0,{\rm
x}}(\xi)|V|^2/\Delta
\end{equation}
with $s = -5/2, -3/2, -1/2, 1/2, 3/2$. The family of lower roots
$F^{(s)}_0(\xi)$ of Eq. (3.9) changing smoothly from $-
\beta_\uparrow^2 - g$ at $\xi = \infty$ to approximately $ - 6/5 - 5g
\beta_\uparrow^2/6$ at $\xi=0$ describe the renormalization of the
Zeeman grid of impurity spin-flipped states due to the kinematic
exchange with LL continuum. To illustrate this dependence we have found
the solution of Eq. (\ref{3.10}) for $g = 0.25$ neglecting
$\varepsilon_{\rm Z}$ and modeling the spin-exciton dispersion by the
function $e(q) = 2 - 2 e^{-q^2/4} I_0(q^2/4)$, which corresponds to the
ideal 2D case.\cite{Lelo80,Bychok81,KH84} (At the same time the
parameter $M_{\rm x}$ may be considered as an empirical value.) The
results of this calculation are presented in Fig. 4.

\begin{figure}[h]\begin{center} \vspace{-9.mm}
\includegraphics*[width=.7\textwidth]{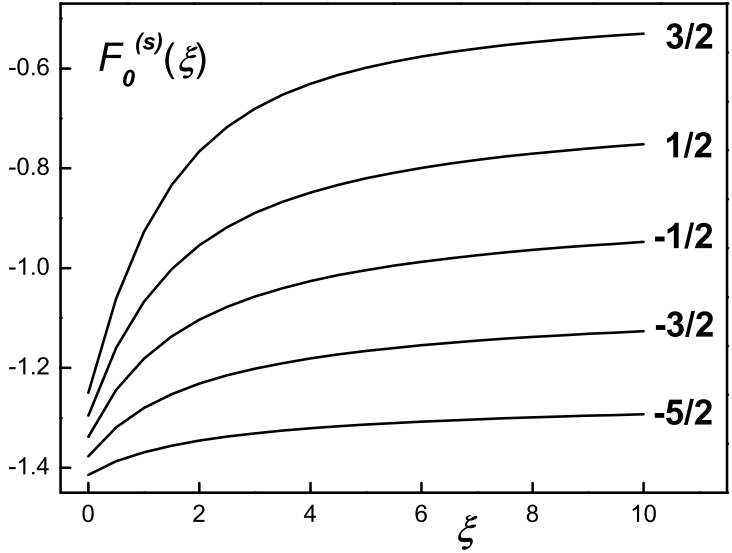}
\end{center}
\vspace{-13.mm} \caption{The lower root of Eq. (\ref{3.9}) with $g =
0.25$. The numbers $s$ are indicated near the curves. See text for
further details.} \label{f.4}
\end{figure}

\subsection{Delocalized impurity-related excitations.}
\label{III.C}

We conclude this section by a brief discussion of the delocalized
states (free spin waves distorted by the resonance magnetic impurity
scattering). These states are described by the functions $f(q)$ with a
divergent norm in the expansion (\ref{24}). Secular equation for these
states cannot be presented in the form (\ref{secular}) but there are
solutions satisfying Eqs. (\ref{equations}) at any energy within the
spin-exciton band, $\varepsilon_{\rm Z} < E < \varepsilon_{\rm Z} +
{\cal E}_{\infty}$. These states are the 'counterparts' of the levels
$E^{(s)}_{\rm x}$ in the spin wave continuum. Let $q_0(E)$ be a root of
equation $\varepsilon_{\rm Z} + {\cal E}_{q_0} = E$. Substituting
\begin{equation}\label{D.1}
f(q) = C\frac{\sqrt{2\pi}}{4q_0} \delta_{|{\bf q}|,q_0} + u(q) (1 -
\delta_{|{\bf q}|,q_0})
\end{equation}
into Eqs. (\ref{equations}) one gets three equations for the
coefficients $D_0$, $C$ and $u(q)$. Turning from summation to
integration and using the rule $\sum_{\bf q}\delta_{|{\bf q}|,q_0} =
{2q_0L}/{\pi}\;$ ($L^2\!=\!2\pi\! N_\phi$, being the 2DEG area) one
finds the coefficient $u(q)$ from the equation $\beta_\uparrow^2(E -
\varepsilon_{\rm Z} - {\cal E}_{q}) u(q) = -
\beta_\downarrow^2N_\phi^{-1/2} h_{nn}(q) (g_i\mu_BB + E)$. Then
equation
\begin{equation}\label{D.2}
C e^{-q_0^2/4} = 1 + \frac{(E + g_i\mu_BB)}{\beta_\uparrow^2}
\left({\beta_\downarrow^2}
\begin{matrix}{{}\quad\displaystyle{\vphantom {\hat A}}_\infty}\\
{\displaystyle \fpint{}{}}\\{{}\!\! \vphantom{\displaystyle {\overline
A}}}^0\end{matrix} \!\!\frac{qdq|h_{nn}|^2}{E-\varepsilon_{\rm Z}
-{\cal E}_{q}}+\frac{{\cal E}_{d,0}-E}{|V|^2}\right)\,
\end{equation}
for the spectrum is derived from Eq. (\ref{equations}) in the
thermodynamic limit ($L,N_\phi\to\infty$). It can be readily seen that
the norm of the function (\ref{D.1}) diverges as $\sum_{\bf
q}|u(q)|^2\sim N_\phi$.

\section{Positive $\displaystyle{\mbox{\large\boldmath $g$}}$$_{\mbox{\scriptsize
2DEG}}$-factor. Pinning of the QHF spin}\label{IV}

Experimentally the magnitude of the $g_{\mbox{\tiny 2DEG}}$ factor
in ${\rm GaAs}/{\rm Al}_x{\rm Ga}_{1-x}$As structures can be altered
gradually by changing pressure or by varying Al content ($x$). It
can be made very small and even change its sign.\cite{ma96} The
value of $g_{\mbox{\tiny 2DEG}}$ may be effectively reduced also due
to optical orientation of nuclear spins changing the electron Zeeman
splitting (Overhauser shift).\cite{ba95,ku99} In this section we
discuss the impurity-related reconstruction of the ground state and
the spectrum of spin-flip excitations at small but positive values
of $g_{\mbox{\tiny 2DEG}}$. It will be shown below that even a
minute amount of magnetic impurities can drastically influence the
QHF state.

Keeping the previous notations, it is now convenient to redirect the
magnetization axis $(\hat z\to -\hat z)$, i.e. to make formal
transformation $g_i \to -g_i$ instead of changing the sign of
$g_{\mbox{\tiny 2DEG}}$. It is clear that at least in the absence of
the $s$-$d$ hybridization the global vacuum state
$|d^5,5/2;|\mbox{vac}\rangle$ serves as the ground state, and all the
spin-flips cost positive energy.  The localized states can still be
found from Eq. (\ref{secular}) with redefined Zeeman energies,
$g_i\mu_BB \to -g_i\mu_BB$ and $\varepsilon_{\rm Z}\to \varepsilon_{\rm
Z}^*$.  The latter parameter actually takes the values
$\varepsilon_{\rm Z}^*=g^*_{\rm 2DEG}\mu_BB\gapprox 0.1\,$K. Making
change $g\to -g$ in Eq. (\ref{3.9}), we denote the lower root of this
new equation  as $F_{\underline {\rm x}}^{(s)}$. This root corresponds
to the energy of the localized spin exciton with changed impurity spin
projection, $\delta S_z^{(d)}\!\approx\!3/2\!-\!s$, where
$s\!=\!\frac{3}{2},\frac{1}{2},-\frac{1}{2},-\frac{3}{2},-\frac{5}{2}$.
The total spin component is $S_z\!=\!-\frac{N_\phi}{2}\!-\!s$ (when
presenting results, we return to the `normal' coordinate system where
$\hat z$ is directed along $\vec B$), and the energy counted off the
global vacuum level is given by formula
\begin{equation}\label{sf-energy}
{\tilde E}_{\underline{\rm x}}^{(s)} = g_i\mu_BB(3/2-s) +
F^{(s)}_{\underline{\rm x}}(\xi)|V|^2/\Delta\,.
\end{equation}
(It should be noted that now the new global vacuum is really below the
old one by the energy $5g_i\mu_BB$.) Functions $F^{(s)}_{\underline{\rm
x}}(\xi)$ are presented in Fig. 5.

\begin{figure}[h]\begin{center} \vspace{-7.mm}
\includegraphics*[width=.7\textwidth]{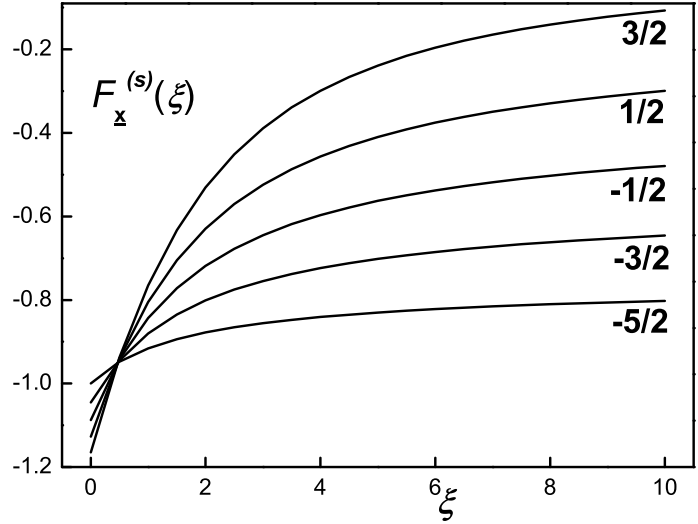}
\end{center}
\vspace{-15.mm} \caption{The lower root of Eq. (\ref{3.9}) with
negative parameter $g$. Calculation is performed for $g\!=\!-0.25$ and
$e(q)\!=\! 2\! -\! 2 e^{-q^2/4} I_0(q^2/4)$. The values of $s$ are
indicated near the curves.} \label{f.5}
\end{figure}

Other roots of the secular equation belong to the continuous spectrum.
These states may be analyzed following the approach described in
Subsection \ref{III.C}. The special `resonance' solution of Eq.
(\ref{D.2}) with $g_i$ substituted for $-g_i$ arises in this case at
$E\!=\!g_i\mu_BB
> \varepsilon_{\rm Z}^*$. Then $u(q) = 0$ and the norm $\langle
X|X\rangle$ diverges  not as $\sim N_\phi$ but as $L\sim N_\phi^{1/2}$
(see discussion in the next section). As a function of $s$, the
delocalized `resonance' states form a set of equidistant levels
\begin{equation}\label{res}
{\tilde E}^{(s)}_{\rm res}=g_i\mu_BB\left(\frac{5}{2}-s\right)
\end{equation}
(again the energy of global vacuum is taken as the reference level).

When looking for the $\underline {\rm x}$-type solutions at
${}\;E\!<\!\varepsilon_{\rm Z}^*\;$ but ${}\;|E|\ll 1/M_{\rm x}\;$, one
may use Eq. (\ref{4.1}). Then one obtains for the localized
spin-exciton energy$\,$ \cite{foot5} ~$E_{\underline{\rm
x}}^{(s)}\!=\!|V|^2F^{(s)}_{\underline{\rm x}}(\xi)/\Delta$ the
following equation
\begin{equation}\label{5.1}
E_{\underline{\rm x}}^{(s)} \approx \varepsilon_{\rm Z}^* -
\frac{1}{\gamma M_{\rm x}}
\exp{\left(\frac{\beta_\uparrow^2}{\beta_\downarrow^2 M_{\rm
x}g_i\mu_BB} - \frac{\Delta} {\beta_\downarrow^2 M_{\rm x}| V
|^2}\right)}\,,
\end{equation}
instead of (\ref{3.8}). Here $s = \frac{3}{2}\,$ has to be taken for
the excitation from the ground state, then $\beta_\uparrow^2 =
\frac{1}{5}\,$ and $\beta_\downarrow \!=\!1$. The exponentially small
energy $E_{\underline{\rm x}}^{(3/2)}$ thus corresponds to formation of
bound spin exciton of large radius. However, for sufficiently small
$\varepsilon_{\rm Z}^*$ (or for a strong enough kinematic exchange) the
energy $E_{\underline{\rm x}}^{(3/2)}$  becomes negative which means an
instability of the global vacuum $|d^5,5/2;|\mbox{vac}\rangle$
considered as the QHF ground state. This instability appears provided
\begin{equation}\label{cond1}
  \xi<\xi_{c1}\,,
\end{equation}
where $\xi_{c1}$ is determined by the equation
\begin{equation}\label{c1}
|V|^2F^{(3/2)}_{\underline{\rm x}}(\xi_{c1})/\Delta+\varepsilon_{\rm
Z}^*=0\,.
\end{equation}

The question arises, whether  the condition (\ref{cond1}) mean the
global reconstruction of the ground state and appearance of a new state
with {\em many} spin excitons bound to the magnetic impurity? To
clarify this point, we discuss the limiting situation where
$\varepsilon_{\rm Z}^*\!\to\!0$ but still $N_{\phi}\varepsilon_{\rm
Z}^*\!\to\!\infty$. Then the ground state at any $\xi$ is no longer the
global vacuum because creation of one spin exciton bound to the
impurity lowers the energy of the system. The corresponding energy gain
compared to the global vacuum is presented as $
|V|^2G_{1}(\xi)/\Delta$. [The subscript `1' corresponds to one bound
exciton; specifically, we have $G_{1}(\xi)\!=\!F_{\underline{\rm
x}}^{(3/2)}$.] To answer the question, one should consider the
situation with $K$ captured spin excitons (then
$S_z\!=\!K\!-\!\frac{N_\phi+5}{2}$) and calculate the proper value
$G_K(\xi)$ at arbitrary $K$. The latter is determined by the
competition between the antiferromagnetic kinematic exchange, which
forces 2DEG spins to reorient in the direction opposite to the impurity
spin, and the Coulomb-exchange energy appearing due to 2DEG
inhomogeneity in a cluster of $K$ spin excitons bound to the impurity.
This inhomogeneity energy is measured in $1/M_{\rm x}$ units.
Calculation of $G_K$ at $K\gtrsim 1$ (but $K\!\not=1\!$) is beyond the
abilities of our present approach but we can consider the case of
$K\!\gg\!1$ and find the conditions under which such a massive pinning
of 2DEG spins in the vicinity of the impurity turns out more
advantageous than binding of single spin exciton (i.e.
$G_\infty\!>\!G_1$).

\subsection{Skyrmionic states created by magnetic impurities}
\label{IV.A}
%\subsection{Skyrmion pinned to magnetic impurity}
%\label{IV.A}

The state with $K\!\gg\!1$ can be described as a collective topological
defect (skyrmion) pinned to a magnetic impurity.\cite{Dick} A smooth
inhomogeneity in the system of spins may be presented as a continuous
rotation in the 3D space. If one characterizes the local direction of
the spins by a unit vector ${\vec n}({\bf r})$ with components $n_x =
\sin{\theta} \cos{\varphi}$, $n_y = \sin{\theta} \sin{\varphi}$, and
$n_z = \cos{\theta}$ ($\varphi$ and $\theta$ are the two first Eulerian
angles) then the conditions $\left.\theta\right|_{{\bf r}=0}=0$ and
$\left.\theta\right|_{{\bf r} = \infty} = \pi$ inevitably result in the
appearance of the topological invariant $q_{\mbox{\tiny T} } = \int
d{\bf r}\rho({\bf r})$ where the density
\begin{equation}\label{rho}
\rho({\bf r}) = \frac{1}{4\pi}{\vec n} \cdot \left(\partial_x {\vec
n}\right) \times \left(\partial_y{\vec n}\right)
\end{equation}
is a vortex characteristics of the spatial twist. The value
$q_{\mbox{\tiny T}}$ has to be an integer nonzero number.\cite{be75}
Its physical meaning is the number of excessive ($q_{\mbox{\tiny
T}}<0$) or deficient ($q_{\mbox{\tiny T}}>0$) electrons in the
system,\cite{so93,fe94,ip97,di02} i.e. $q_{\mbox{\tiny T}}=N_\phi - N$.
In a perfect 2DEG and at nearly zero Zeeman gap ($\varepsilon_{\rm
Z}^*\!\to\!0$) such a weakly inhomogeneous skyrmion state has the
energy
\begin{equation}\label{E_sk}
{\cal E}_{\rm sk} = \frac{3}{4}{\cal E}_{\infty}q_{\mbox{\tiny T}} +
\frac{1}{2M_{\rm x}} \left(\left|q_{\mbox{\tiny
T}}\right|-q_{\mbox{\tiny T}}\right)\,.
\end{equation}
This result is valid within the single Landau level approximation (see,
e.g., Ref. \onlinecite{di02}). It is enough to consider the case
$q_{\mbox{\tiny T}}\!=\!\pm 1$, because any state with $|q_{\mbox{\tiny
T}}|\!>1$ is merely a combination of `singly-charged' skyrmions. Due to
the hybridization with the impurity the skyrmionic state gains a
negative kinematic exchange energy. The latter has to be taken into
account in combination with the Coulomb-exchange energy (\ref{E_sk})
and with the finite positive Zeeman energy at $g_{\mbox{\tiny
2DEG}}^*>0$
\begin{equation}\label{K}
E_{\rm Z} = \varepsilon_{\rm Z}^*K, \quad\mbox{where}\quad
K=\frac{1}{4\pi l_B^2} \int (1 + \cos{\theta})d{\bf r}
\end{equation}
(in the clean 2DEG the skyrmion energy is given by ${\cal E}_{\rm
sk}\!+\!E_{\rm Z}$).

One can conclude from symmetry considerations that the impurity is
located at the center of the topological defect. Then additional
pinning energy may be found by means of the conventional energy
minimization procedure where the Euler angles are used as variational
parameters. This energy is a difference between the energy of the
global vacuum state with a distant skyrmion and the ground state energy
calculated in the presence of magnetic impurity at the center of the
topological defect (cf. Ref. \onlinecite{Dick} where similar procedure
was elaborated in the limit of potential impurity scattering). Namely,
to calculate the contribution of magnetic impurity at $K\!\gg\!1$, one
should consider a domain around impurity which is small in comparison
with a characteristic area of the skyrmion, but contains a large enough
number of  spin-flipped LL electrons involved in the formation of
pinned topological defect. Then the situation becomes similar to that
considered in Sec. III:  $s$-$d$ hybridization of the impurity electron
with the $m\!=\!0$ electron in this domain generates the kinematic
exchange in accordance with Fig. \ref{f.1}, and leads to reconstruction
of the spectrum in accordance with Eq. (\ref{secular}). The shift of
the energy with respect to the global vacuum is given by the value
$5g_i\mu_BB\!+\!{\tilde E}_0^{(-5/2)}$, where ${\tilde E}_0^{(-5/2)}$
is determined by Eq. (\ref{3.10}).\cite{foot6} Hence we obtain that the
pining energy is $ E_{\rm sk,pin}\! =\!-g_i\mu_BB
-F^{(-5/2)}_{0}(\xi)|V|^2/\Delta$, where $F^{(-5/2)}_0$ is shown in
Fig. 4. In the limit of strong pining ($E_{\rm sk,pin} \gg {\cal
E}_{\rm sk}$) and `frozen' impurity spin ($g\gg 1$) this result agrees
with the pinning energy found earlier.\cite{Dick}

The energy $E_{\rm sk,pin}$ is calculated in the leading approximation,
which does not depend on the charge $q_{\mbox{\tiny T}}$. However, it
is instructive to obtain the correction related to the inhomogeneity of
the texture. It is known\cite{ip97,di02} that the density (\ref{rho})
may be interpreted in terms of effective magnetic field appearing in
the Schr\"odinger equation due to this inhomogeneity. One may introduce
the renormalized magnetic length $l_B\to {\tilde l_B}$ as
\begin{equation}\label{l_B}
\frac{1}{{\tilde l_B}^2} = \frac{1}{l_B^2} - 2\pi\rho({\bf r}).
\end{equation}
Taking into account that $|V|^2/\Delta\sim 1/l_B^2$ and $\xi\sim l_B$
and rewriting Eqs. (\ref{rho}) and (\ref{l_B}) in terms of the Euler
angles$,$\cite{be75} one finds the correction to pinning energy due to
the finite radius $R^*$ of the skyrmion core (see Ref.
\onlinecite{Dick} for a detailed calculation). The corrected energy is
determined by the value $\rho(0)$ and has the form
\begin{equation}\label{E_sk,pin}
E_{\rm sk,pin}^{(\mbox{\footnotesize\it q}_{\mbox{\tiny T} })} =
-g_i\mu_BB-\frac{|V|^2}{\Delta} \left[F_0^{(-5/2)}(\xi)-q_{\mbox{\tiny
T}}\!\left(\frac{l_B}{R^*}\!\right)^2
\left(2F_0^{(-5/2)}-\xi\frac{dF_0^{(-5/2)}} {d\xi}\right)\right],\quad
\ q_{\mbox{\tiny T}}=\pm 1\,.
\end{equation}
It is assumed here that $g\ll 1$.

The skyrmion core radius $R^*$ is found by considering the competition
between the Zeeman energy (\ref{K}) and the energy of Coulomb
repulsion.\cite{so93,by98} Generally speaking, in our case in order to
find $R^*$ we should include the energy $E_{\rm sk,pin}$ in the
minimization procedure. However, this correction only insignificantly
influences the result due to the condition $R^*\gg l_B$ and because of
the rather strong $e$-$e$ interaction resulting in the skyrmion
formation. Using the realistic estimate for the kinematic exchange
energy $|V|^2/ \Delta \lapprox E_{\rm C}$ the minimization yields the
same formula
\begin{equation}\label{K1}
E_{\rm Z}=\frac{\varepsilon^*_{\rm Z}}{2}(R^*/l_B)^2 \ln{(l_B^2E_{\rm
C}/\varepsilon^*_{\rm Z}{R^*}^2)}
\end{equation}
as in the case of `free' skyrmions$\,$\cite{by98}, where ${R^*}^3 =
{9\pi^2l_B^2e^2}/ \left[{64\varepsilon^*_{\rm Z} \kappa \ln{(E_{\rm C}/
\varepsilon_{\rm Z}^*)}}\right]$. The number of spin-flipped electrons
turns out to be rather large
\begin{equation}\label{K2}
K = \frac{1}{96} \left( \frac{9\pi^2e^2}{\kappa\varepsilon_{\rm
Z}^*l_B}\right)^{2/3} \left[\ln{(E_{\rm C}/\varepsilon_{\rm
Z}^*)}\right]^{1/3} \sim 10-20 \quad(\mbox{if}\;\; \varepsilon^*_{\rm
Z}\sim 0.1\,\mbox{K})\,.
\end{equation}

We first consider the regime where there is no skyrmions in the clean
system but these collective excitations could be created due to strong
enough kinematic exchange interaction between the LL electrons and
magnetic impurities. This is the situation where the inequality
(\ref{cond1}) is valid, and besides the condition $\left|N \!- \!N_\phi
\right|\!\ll\! N_i$ is realized, where $N_i$ is the number of
impurities. The electroneutrality of the system requires that the
topological defects are created as skyrmion-antiskyrmion pairs. Two
impurities are able to create a skyrmion-antiskyrmion pair provided the
pinning energy $E_{\rm sk,pin}^{(+)} \!+\! E_{\rm sk,pin}^{(-)}$
exceeds the energy increase due to the skyrmion-antiskyrmion gap. The
latter in accordance with Eq. (\ref{E_sk}) includes the
Coulomb-exchange part equal to $M_{\rm x}^{-1}$ and twice the Zeeman
energy [Eqs. (\ref{K}) and (\ref{K2})]. In addition, the energy of a
skyrmion and an antiskyrmion pinned by two neighbouring magnetic
impurities has to be lower than the double energy of a pinned spin
exciton. Thus the condition $G_\infty\!<\!G_1$  for creation of a
pinned skyrmion-antiskyrmion pair can be rewritten in the form
\begin{equation}\label{cond2}
\xi\!<\!\xi_{c,\infty}\,,
\end{equation}
where the critical value $\xi_{c,\infty}$ can be obtained with the help
of Eq. (\ref{E_sk,pin}):
\begin{equation}\label{c,infty}
g+F^{(-5/2)}_0(\xi_{c,\infty})+ \xi_{c,\infty}/2 + E_{\rm
Z}\Delta/|V|^2=F_{\underline{\rm x}}^{(3/2)}(\xi_{c,\infty})\,.
\end{equation}
Under the condition (\ref{cond2}) an impurity  acquires the
localized magnetic moment $K \sim B^{-1/3}$ antiparallel to its
own moment and exceeding it (when, e.g. $K > 5/2$ in the GaAs:Mn
case).

\subsection{Phase diagram of QHF ground state at
\mbox{\boldmath $g_{\mbox{\tiny 2DEG}}^*>0$} } \label{IV.B}

There are two critical transitions in our problem: first, the global
vacuum is destroyed when $\xi$ becomes less than $\xi_{c1}$ and single
spin-flip exciton appears (this state may be characterized as a `local
pinning'); second, the massive pinning of 2DEG spins takes place when
$\xi$ reaches the value $\xi_{c,\infty}$. However, this scenario is
somewhat changed if one takes into account finite ratios $N_i/N_\phi$.
Indeed, up to this point we have supposed that the Zeeman energy
$\varepsilon^*_{\rm Z}N_\phi$ corresponding to the `global flip' of all
2DEG spins is larger than any contribution to the QHF energy due to the
magnetic impurities. This global spin-flip actually represents the spin
configuration treated as the ground state in the previous section. When
counted from the global vacuum, its energy per one impurity is $E_{{\rm
s}\mbox{\tiny -}{\rm f}}=g_i\mu_BB+N_\phi\varepsilon^*_{\rm
Z}/N_i\!+\!|V|^2F^{(-5/2)}_0(\xi)/\Delta$. Negative $E_{{\rm
s}\mbox{\tiny -}{\rm f}}$ means that available magnetic impurities are
able to  polarize completely all 2DEG electrons even at positive
$g_{\mbox{\tiny 2DEG}}^*$. In agreement with the above discussion, one
can conclude that such a complete polarization takes place when
\begin{equation}\label{pinning}
E_{{\rm s}\mbox{\tiny -}{\rm f}}(\xi)<E_{\rm min}(\xi)\,,
\end{equation}
where $E_{{\rm min}}=\min\{0,~~ |V|^2F_{\underline{\rm
x}}^{(3/2)}(\xi)/\Delta\!+\!\varepsilon_Z^*,~~
g_i\mu_BB\!+\!|V|^2F^{(-5/2)}_0(\xi)/\Delta\!+\!\frac{1}{2}M_{\rm
x}^{-1}\!+\! E_{\rm Z}\}$.

\vspace{-5.mm}
\begin{figure}[h]\begin{center} \vspace{-0.mm}
\includegraphics*[width=7.\textwidth]{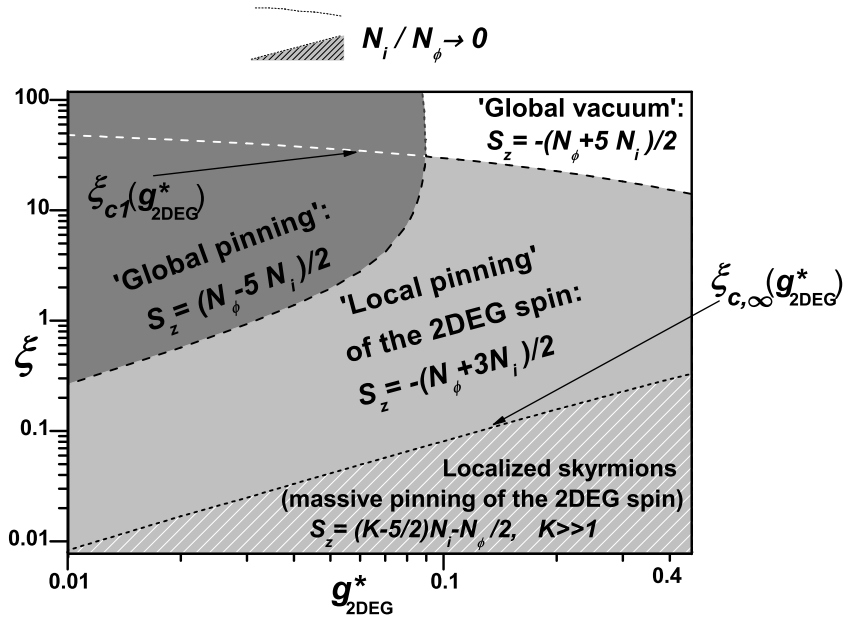}
\end{center}
\vspace{-15.mm} \caption{Phase diagram illustrating the reconstruction
of the QHF ground state at $g_{\mbox{\tiny 2DEG}}^*\!>\! 0$ for two
cases: isolated impurity $N_i/N_\phi\to 0$ (see elucidating legend
above the main picture) and finite impurity concentration
$N_i/N_\phi=0.01$. The calculation was carried out for the Zeeman
parameters $g\!=\! g_i\mu_BB\Delta/|V|^2 \!=\!0.25$ and
$\varepsilon_{\rm Z}^* = 0.05(g_{\mbox{\tiny
2DEG}}^*/0.44)|V|^2/\Delta$ and for the spin exciton dispersion equal
to $\;\xi e(q)=0.05(g_{\mbox{\tiny 2DEG}}^*/0.44)+2\xi\left[1\!-\!
e^{-q^2/4} I_0(q^2/4) \right]$ in $|V|^2/\Delta$ units. Comments in the
figure refer to the $N_i/N_\phi= 0.01$ case. See text for other
details.} \label{f.6}
\end{figure}

The phase diagram of our system at zero temperature is determined by
the interplay between Zeeman splitting, Coulomb interaction and {
kinematic} impurity exchange energy, and controlled by the impurity
concentration. These factors are characterized by the dimensionless
parameters $g_{\mbox{\tiny 2DEG}}^*$, $\xi$ and $N_i/N_\phi$. One can
construct this diagram by employing the inequalities (\ref{cond1}),
(\ref{cond2}) and (\ref{pinning}). The results for both cases of
infinitely small and finite ratio $N_i/N_\phi$ are presented in Fig.
\ref{f.6} in the $(g_{\mbox{\tiny 2DEG}}^*,\xi)$ coordinates. We
expressed the $e$-$e$ interaction values entering the skyrmion Zeeman
energy [Eqs. (\ref{K1}) and (\ref{K2})] in terms of the parameter
$M_{\rm x}$: $E_{\rm C} = M_{\rm x}^{-1}$, $e^2/\kappa l_B = M_{\rm
x}^{-1} (8/\pi)^{1/2}$. The phase diagram for the $N_i/N_\phi \to 0$
case is explicated by the legend above the main picture.  The
$\xi\!=\!\xi_{c1}(g_{\mbox{\tiny 2DEG}}^*)$ curve in Fig. \ref{f.6}
separates states with unbroken global vacuum (the area above this line)
and states of `local pinning' where each impurity is dressed by one
bound spin-exciton.  The dotted line [$\xi_{c,\infty}(g_{\mbox{\tiny
2DEG}}^*)$ curve] separates the state with `local pinning' and the
state of massive spin reversal ($K\gg 1$) determined by the pinned
skyrmions (hatched area below this line).

In the more realistic case of $N_i/N_\phi\!=\!0.01$, the curves
$\xi_{c1}(g_{\mbox{\tiny 2DEG}}^*)$ and $\xi_{c,\infty}(g_{\mbox{\tiny
2DEG}}^*)$ formally remain the same since the parameter $N_i/N_\phi$
does not enter Eqs. (\ref{c1}) and (\ref{c,infty}). However, in this
case essential part of the $(g_{\mbox{\tiny 2DEG}}^*,\xi)$ area belongs
to states where the 2DEG spins are globally polarized in the ${\vec B}$
direction in spite of positive $g_{\mbox{\tiny 2DEG}}^*$.
 This area filled by dark-grey presents solutions of
inequality (\ref{pinning}). Unbroken global vacuum occupies only the
blank sector in the upper right corner of the phase diagram. At large
$\xi$ but fixed $|V|^2/\Delta$, the line separating the blank and
dark-grey sectors tends to $g_{\mbox{\tiny 2DEG}}^*\!=\!0.088$, which
corresponds to value $\varepsilon^*_{\rm
Z}\!=\!N_i|V|^2\!/N_\phi\Delta$ being the result of the $E_{\rm
s-f}(\xi\!\to\!\infty)\!=\!0$ equation. At the same time if the
$\xi\!\to\!\infty$ limit is realized owing to vanishing $V$, then both
systems of the impurities and of the 2DEG become independent and at any
positive $g_{\mbox{\tiny 2DEG}}^*$ the global vacuum presents certainly
the ground state. The light-grey area below the
$\xi\!=\!\xi_{c1}(g_{\mbox{\tiny 2DEG}}^*)$ line but above the dotted
line corresponds to the singly spin-flip states with one exciton bound
to impurity. The hatched light-grey domain below the dotted line
corresponds to the state with the localized skyrmions created by strong
kinematic exchange [Eq. (\ref{cond2})]. In our specific case of the
$N_i/N_\phi=0.01$ ratio the dark-grey sector is not contiguous to this
skyrmionic region. The total QHF spin $S_z$ in various states of the
phase diagram is indicated in the picture.

Now we discuss the regime where free skyrmions are already available in
the system because the number of electrons well deviates from the
quantum flux number. Namely,  we consider that $\left|N \!- \!N_\phi
\right|\!>\! N_i$ (although still $\left|N \!- \!N_\phi
\right|\!\ll\!N_\phi$). In this case `excessive' skyrmions may be bound
to an impurity. The result depends on the QHF phase. In the globally
pinned phase (dark-grey area) the binding is impossible since the
effective interaction between the impurity and the skyrmion is
repulsive. In the state of local pinning (light-grey unhatched domain)
the binding also does not occur. Indeed, the binding energy would be
equal to $E_{\rm sk,pin}$ (\ref{E_sk,pin}) but due to the condition
(\ref{cond1}) this value is smaller than the spin exciton
delocalization energy $-|V|^2F_{\underline{\rm x}}^{(3/2)}/\Delta$. At
the global vacuum (blank sector) the binding takes place and the
binding energy is equal to the pinning energy (\ref{E_sk,pin}).
Certainly, the binding takes place in the skyrmionic ground state
(light-grey hatched sector). However, in contrast to the
$|N\!-\!N_\phi|\!<\!N_i\,$ case, now all $N_i$ impurities bind
skyrmions of the same charge $q_{\mbox{\tiny T}}$, where
$q_{\mbox{\tiny T}}\!=\pm 1$ if correspondingly $N\lg N_\phi$.

To conclude this section, it is worthy to remind that we have only
considered the situation where the $g_{\mbox{\tiny 2DEG}}^*\!>\!0$
ground state is realized in the most symmetric phases when the pinned
spin $K$ is equal to 0,$\;$ 1 or $K\!\gg\!\infty$. As it has been seen,
there are only two critical parameters $\xi_{c1}$ and $\xi_{c,\infty}$
in this case. However, one might suppose that transition from the local
pinning ($K\!=\!1$) to the skyrmionic phase of massive pinning  would
proceed more smoothly with diminishing parameter $\xi$. Namely, below
the $\xi_{c1}(g_{\mbox{\tiny 2DEG}}^*)$ curve there should be critical
value $\xi=\xi_{c2}(g_{\mbox{\tiny 2DEG}}^*)$ at which the transition
$K\!=\!1\to K\!=\!2$ occurs. This value would be the root of equation
$G_1(\xi_{c2})\!=\!G_2(\xi_{c2})$. The next critical point would
correspond to the $K\!=\!2\to K\!=\!3$ transition and so on. This
sequence of values $\xi_{c1}\!>\!\xi_{c2}\!>\!...\xi_{cK}\!>\!...$
where $G_{K\!-\!1}(\xi_{cK})\!=\!G_K(\xi_{cK})$ should condense in the
vicinity of $\xi_{c,\infty}$. Actually this means that the light-grey
unhatched domain in Fig. \ref{f.6} would present not only the singly
spin-flip 2DEG state but a set of states with $K=1,2,3,...$
spin-excitons localized at the impurity where $K$ is growing with
diminishing $\xi$. In reality, for a finite $\varepsilon_{\rm Z}^*$,
$K$ reaches the value given by Eq. (\ref{K}) at
$\xi\!=\!\xi_{c,\infty}$. Such a `stratification' of the light-grey
unhatched sector would be the only qualitative change of the phase
picture of Fig. \ref{f.6}. Quantitative changes would be presented by
appropriate shifts of the $\xi_{c,\infty}$ curve and of the boundary
between the dark-grey and light-grey areas. However, it is clear that
these shifts would not be significant. The corresponding crossover
parameters $\xi$ would at least remain of the same order as the ones
calculated with the help of Eqs. (\ref{c,infty}) and (\ref{pinning}).

\section{Discussion}

We have found that the interplay between the kinematic impurity
exchange and the Coulomb interaction in 2DEG results in the appearance
of bound exciton states and in the renormalizaton of impurity spin
states, including the reconstruction of the QHF ground state at
$g^*_{\rm \tiny 2DEG} > 0$.

\begin{figure}[h]\begin{center} \vspace{-5.mm}
\includegraphics*[width=.7\textwidth]{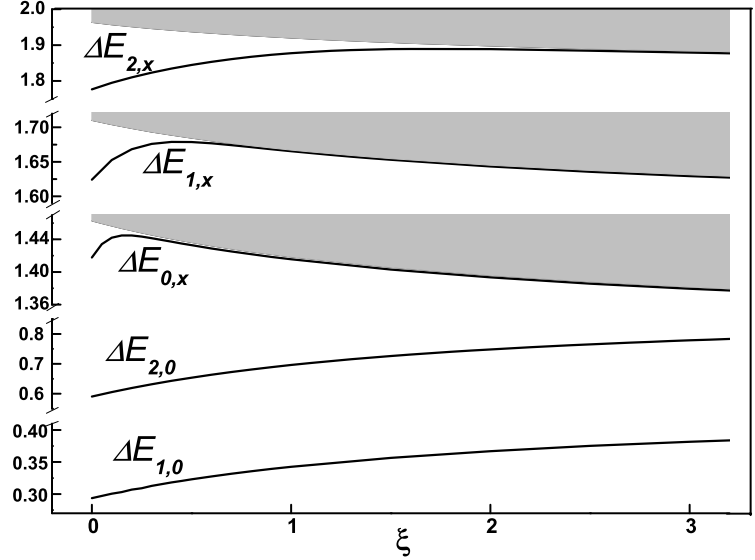}
\end{center}
\vspace{-16.mm} \caption{The case of $g_{\mbox{\tiny 2DEG}} < 0$.
Energies of the excitations measured from the ground state are plotted
in units of $|V|^2/\Delta$ as functions of $\xi=\Delta/M_{\rm x}|V|^2$.
Filled areas show energies of the delocalized spin excitons. See text
for details. } \label{f.7} \vspace{-1.5mm}
\end{figure}

Among the available experimental techniques, the inelastic light
scattering (ILS) method seems to be the most useful method for
experimental study of the 2DEG spectra (see Refs.
\onlinecite{pi92,va06,pi} and references therein). However, this tool
has some special features, and it is helpful to discuss our results
from this point of view. Let us first consider the $g_{\mbox{\tiny
2DEG}} \!<\! 0$ case. When measuring the energy from the ground state
level ${\tilde E}_{0}^{(-5/2)}$ , where the impurity has the maximum
spin projection [see Eq. (\ref{3.10})], one obtains ten levels of the
localized excitations $\Delta E$ related to the spin changes $\delta
S_z=0,1,...5$
\begin{equation}\label{levels}
\Delta E_{\delta S_z,t} = g_i\mu_BB \delta S_z + (|V|^2/\Delta) \left[
F^{(s)}_{t}(\xi) - F_0^{(-5/2)}(\xi)\right]\quad (t=0,\mbox{x}),
\end{equation}
where $s = \delta S_z - 5/2$, and the index $t$ labels the type of the
excited state [in Eq. (\ref{levels}) it is considered that
$F^{(5/2)}_{t}\!\equiv\!0$]. Within the scope of the experiment where
only the $|\delta S_z|\le 2$ excitations seem to be observable as ILS
peaks, we plot in Fig. \ref{f.7}. these five levels calculated with the
help of Eq. (\ref{3.9}) for the parameters $g = 0.25$ and
$\varepsilon_{\rm Z} = 0.05|V|^2/ \Delta$. This calculation is done for
the sake of demonstration with the function ${\cal E}_q = {2}{M_{\rm
x}}^{-1} \left[1 - e^{-q^2/4} I_0(q^2/4) \right]$ and the fitting
parameter $M_{\rm x}$ used to describe the spin-wave dispersion. In the
available wide quantum wells the inverse spin-exciton mass is
relatively small.\cite{gallais} Hence the values $\xi < 1$ seem to be
experimentally relevant, and the evolution of non-equidistant
excitations $\Delta E_{\delta S_z,{\rm x}}$ as a function of $\xi$ (and
therefore of $B$) should be observable in this interval.

{ The non-localized states discussed in Section \ref{III.C} actually
present a transformation of the x-type excitations when the spin
exciton is detached from the impurity and falls in the spin-wave
continuum. The bottoms of continuous bands are shown as filled areas in
Fig. \ref{f.7}. The band edges are higher than the $\Delta E_{\delta
S_z,{\rm x}}$ curves by the quantity
$-\!\left(|V|^2/\Delta\right)F^{(\delta S_z\!-\!\frac{5}{2})}_{{\rm
x}}(\xi)$ [see Eqs. (\ref{3.8}) and (\ref{3.10})], and therefore the
latter may be treated as spin-exciton binding energy. However, it seems
to be difficult to observe these states in the ILS spectra because of
comparatively small oscillator strengths, specifically, due to
divergence of the envelope function $f(q)$.

Similar ILS picture should also take place for $g_{\mbox{\tiny 2DEG}}^*
\!>\! 0$ in the phase of the 2DEG global pinning (dark-grey area in
Fig. \ref{f.6}). In the skyrmionic phase (light-grey hatched sector)
there are intra-impurity ILS transitions determined by Eq.
(\ref{levels}). Besides, two new types of resonances are expected: the
first is the skyrmion delocalization with $\delta S_z\!=\!0$ and with
excitation energy equal to $E_{\rm sk,pin}$ (\ref{E_sk,pin}); another
one is transition $\delta S_z\!=\!-1$ where the delocalized skyrmion
leaves the impurity with the bound spin exciton. In the latter case the
transition energy is $E_{\rm sk,pin}\!-\!|V|^2F_{\underline{\rm
x}}^{(3/2)}\!+\!\varepsilon_{\rm Z}^*$.

In the global vacuum and local pinning states (blank and light-grey
unhatched domains) the ILS spectrum is determined by transitions
between levels (\ref{sf-energy}) and (\ref{res}), so that, e.g., the
ILS transitions to the localized states from the global vacuum are
determined by the energies $\Delta E_{\delta S_z}=\varepsilon_{\rm
Z}^*+g_i\mu_BB(\delta S_z\!-\!1)+|V|^2F_{\underline{\rm
x}}^{(5/2\!-\!\delta S_z)}/\Delta$ and correspond to nonzero spin
change $\delta S_z\!=\!1,2,(3,4,5)$. At the same quantum numbers
$\delta S_z$ there should also be resonance features related to the
impurity spin rotation, which cost the energy $\Delta E_{\delta
S_z,{\rm res}}\!=\!g_i\mu_BB\delta S_z$. These resonances are in
fact transitions to the continuous spectrum which should be
noticeable on the background of free spin waves contribution
($\varepsilon_{\rm Z}^*\! < \Delta E\! < \!{\cal E}_\infty$) [see
the comment above Eq. (\ref{res})].\cite{foot7} The ILS spectrum of
excitations from the local-pinning ground state is presented: first,
by the $\delta S_z\!=\!-1$ transition to the global vacuum (this
energy is equal to $\Delta E_{-1}=-|V|^2F_{\underline{\rm
x}}^{(3/2)}/\Delta-\varepsilon_{\rm Z}^*$); second, by the $\delta
S_z=1,2,(3,4)$ transitions to the localized spin-flip states with
energies $\Delta E_{\delta S_z}=g_i\mu_BB\delta S_z+|V|^2
  \left[F_{\underline{\rm x}}^{(3/2\!-\!\delta S_z)}\!-\!F_{\underline{\rm
x}}^{(3/2)}\right]/\Delta\,; $ and third, by the $\delta
S_z=0,1,2,(3,4)$ transitions to the resonance states in the continuous
spectrum with transition energies $ \Delta E_{\delta S_z,{\rm
res}}\!=\!g_i\mu_BB(\delta S_z\!+\!1)-|V|^2F_{\underline{\rm
x}}^{(3/2)}/\Delta-\varepsilon_{\rm Z}^*\,. $

Finally, we note that currently there are several possibilities for the
experimental study of skyrmion-like textures (e.g., see Refs.
\onlinecite{ma96,ba95,ku99} and \onlinecite{osscrp98}). However, for
any method one of the most serious obstacles impeding observation of
spin-flip phases is the very narrow interval in the vicinity of the
$g_{\mbox{\tiny 2DEG}}^*\!=\!0$ factor where the spin-flip
reconstruction of the ground state or skyrmion-like excitations are
possible. From this point of view the minor magnetic doping would
become an additional fine tuning tool allowing to change the balance
between $E_{\rm Z}$ and $E_{\rm C}$ and to influence the skyrmion
formation.

\section{acknowledgments}

{ The work was mainly done during the authors' stay at MPIPKS, Dresden.
S.D. thanks for the hospitality the Grenoble High Magnetic Field
Laboratory and the Abdus Salam International Center of Theoretical
Physics (Trieste) where part of this work was also carried out and
acknowledges support of the Russian Foundation for Basic Research. V.F.
and K.K. acknowledge partial support of Israeli Science Foundation,
grant No. 0603214212.}

\appendix
%\vspace{5mm}

\section{EXCITONIC REPRESENTATION}
\label{A}

The excitonic representation is a convenient tool for a description of
electron-hole states in a 2DEG multiply degenerate in $m$. When acting
on the vacuum state $|\mbox{vac}\rangle$ (in our case this vacuum is
defined in Subsection \ref{II.A}), the exciton-creation operators form
a system of basis states diagonalizing the Hamiltonian including a
considerable part of the Coulomb interaction. Due to translational
invariance of a clean 2DEG these exciton states are classified by the
2D momentum ${\bf q}$ and the degeneracy turns out to be lifted. The
exciton operators for a single LL were first introduced in Refs.
\onlinecite{dz83-84}. The commutation rules for the same case of single
LL were found in Ref. \onlinecite{by87} (see also Ref.
\onlinecite{di05} and references therein).

Unlike previous papers, where the ER technique was developed for the
Landau gauge, {we use the symmetric gauge for bare one-electron
states}. In the {\em symmetric gauge} the spin-exciton creation
operator is expressed in terms of the $a_m$ and $b_m$ Fermi operators,
\begin{equation}\label{52}
{\cal Q}^\dag_{\bf q} = N_\phi^{-1/2}\!{}\!\sum_{m,m' = 0}^{N_\phi -
1} h_{mm'}^*({\bf q}){ b}^\dag_{m - n}{a}_{m'\!-\!n}\,,
\end{equation}
(cf. the definition based on the Landau
gauge$\;$\cite{di05,di02,dz83-84,Dick}). In this expression
\begin{equation}\label{51}
h_{mk}({\bf q}) = \left({m!}/{k!}\right)^{1/2}(q_-)^{k - m} L^{k -
m}_{m}(q^2/2) e^{-q^2/4}
\end{equation}
are the building block functions used in the ER technique, $q_-\!= {i}
q e^{-i\varphi}/ {\sqrt{2}} \equiv {i}(q_x-iq_y)/{\sqrt{2}},\;$ and
$L_m^{k - m}$ are the Laguerre polynomials. Here and below all lengths
are measured in the magnetic length $l_B\!=\!1$ units. The spin-exciton
states are orthogonal and normalized,
\begin{equation}\label{ortho}
 \langle\mbox{vac}| {\cal Q}_{{\bf q}_1}
 {\cal Q}^\dag_{{\bf q}_2}|\mbox{vac}\rangle=\delta_{{\bf
 q}_1,{\bf q}_2}.
\end{equation}
The operators (\ref{52}) together with the intra-sublevel operators
\begin{eqnarray}\label{7}
{\cal A}^\dag_{\bf q}& = & \frac{1}{N_\phi} \sum_{m,m'=0}^{N_\phi -
1} h_{mm'}^*({\bf q}){ a}^\dag_{m - n}{ a}_{m' - n},\quad \mbox{and}
\quad{\cal B}^\dag_{\bf  q} = \frac{1}{N_\phi} \sum_{m,m'=0}^{N_\phi
- 1} h_{mm'}^*({\bf q}){ b}^\dag_{m - n}{b}_{m' - n}\,.
\end{eqnarray}
form a closed Lie algebra. In order to check it we first obtain the
commutation relations
\begin{equation}\label{I.1}
[{\cal Q}_{\bf q}^\dag,{ a}^\dag_m] = {N}_\phi^{-1/2}
\sum_{k=0}^{N_\phi - 1} h^*_{m + n,k}({\bf q}){  b}^\dag_{k - n},\quad
[{\cal Q}_{\bf q}^\dag,{  b}_m] = -{N}_\phi^{-1/2} \sum_{k=0}^{N_\phi -
1} h^*_{k,m + n}({\bf q}){  a}_{k - n},
\end{equation}
\begin{equation}\label{I.2}
[{\cal A}_{\bf q},{  a}_m]=-\frac{1}{{N}_\phi}
\sum_{k=0}^{N_\phi\!-\!1} h_{m\!+\!n,k}({\bf q}){ a}_{k\!-\!n},\quad
[{\cal B}_{\bf q},{ b}^\dag_m]=\frac{1}{{N}_\phi}\sum_{k=0}^{N_\phi
- 1} h_{k,m + n}({\bf q}){  b}^\dag_{k - n},
\end{equation}
and
\begin{equation}\label{I.3}
[{\cal Q}_{\bf q},{  a}^{\dag}_m]=[{\cal Q}_{\bf q},{ b}_m]=[{\cal
A}_{\bf q},{  b}_m] = [{\cal A}_{\bf q},{ b}_m^\dag]=[{\cal B}_{\bf
q},{  a}_m]=[{\cal B}_{\bf q},{ a}_m^\dag]\equiv 0\,.
\end{equation}

As a result we see that operators (\ref{52}) and (\ref{7}) really form
a closed algebra with the commutation
relations$\,$\cite{di05,di02,by87}
\begin{equation}\label{I.4}
\begin{array}{l}
\displaystyle{\left[{\cal Q}_{{\bf q_1}}, {\cal Q}_{{\bf
q_2}}^{+}\right]= e^{i({\bf q}_1\times{\bf q}_2)_z/2}{\cal A}_{\bf
q_1\!- \!q_2}- e^{-i({\bf q}_1\times{\bf q}_2)_z/2}{\cal B}_{\bf
q_1\!-
\!q_2},}\\
\displaystyle{e^{-i({\bf q}_1\times{\bf q}_2)_z/2}[{\cal
A}^\dag_{{\bf q}_1}, {\cal Q}^\dag_{{\bf q}_2}]= -e^{i({\bf
q}_1\times{\bf q}_2)_z/2}[{\cal B}^\dag_{{\bf q}_1}, {\cal
Q}^\dag_{{\bf q}_2}]=
-{N}_\phi^{-1} {\cal Q}^\dag_{{\bf q}_1\!+\!{\bf q}_2},}\\
\displaystyle{[{\cal A}^\dag_{{\bf q}_1}, {\cal A}^\dag_{{\bf
q}_2}]=-\frac{2i}{N_\phi}\sin{[({\bf q}_1\times{\bf q}_2)_z/2]}
{\cal A}^\dag_{{\bf q}_1\!+\!{\bf q}_2},}\\
\displaystyle{[{\cal B}^\dag_{{\bf q}_1}, {\cal B}^\dag_{{\bf
q}_2}]=-\frac{2i}{N_\phi}\sin{[({\bf q}_1\times{\bf q}_2)_z/2]} {\cal B}^\dag_{{\bf
q}_1\!+\!{\bf q}_2}}\,.
\end{array}
\end{equation}

Acting on the vacuum state the intra-sublevel operators result in
\begin{equation}\label{I.5}
{\cal A}^\dag_{{\bf q}}|\mbox{vac}\rangle=\delta_{{\bf
q},0},\quad\mbox{and}\quad {\cal B}^\dag_{{\bf
q}}|\mbox{vac}\rangle\equiv 0\,.
\end{equation}

The excitonic basis ${\cal Q}^\dag_{\bf q}|\mbox{vac}\rangle$ determine
the set of eigenstates of a clean 2DEG,
\begin{equation}\label{I.6}
\left[({\hat H}_1^{(s)}+{\hat H}_{s\!-\!s}),{\cal Q}^\dag_{\bf
q}\right]|\mbox{vac}\rangle=\left(\varepsilon_{\rm Z}+ {\cal
E}_q\right){\cal Q}^\dag_{\bf q}|\mbox{vac}\rangle\,.
\end{equation}
Here ${\cal E}_{q}$ stands for the Coulomb energy of the free spin wave
defined by the equation$\,$\cite{Lelo80,Bychok81,KH84}
\begin{equation}\label{I.7}
{\cal E}_q = \frac{1}{N_\phi}\sum_{\bf p} {W}_{ss}(p) \left[1 -
e^{i({\bf p} \times{\bf q})_z}\right]\equiv \int_0^{\infty} d pp
{v}_{ss}(p) \left[ h_{nn}(p)\right]^2\left[1 - J_0(pq)\right]\,,
\end{equation}
$J_0(pq)$ is the Bessel function.

The Coulomb vertices in the Hamiltonian (\ref{H_ss}), (\ref{H_sd}) are
given by the equations
\begin{equation}\label{2.7}
W_{ss}({ q})={v}_{ss}({q})[h_{nn}({q})]^2,\qquad
W_{sd}({q})={v}_{sd}({q})h_{nn}({q}),
\end{equation}
where $2\pi{v}_{ss}({\bf q})$ and $2\pi{v}_{sd}({\bf q})$ are the 2D
Fourier transforms of the average $s$-$s$ and $s$-$d$ interaction
potentials. One can present them as$\,$\cite{an82}
\begin{eqnarray}\label{coul}
&&{v}_{ss}({ q})=\frac{e^2}{\kappa l_Bq}\int\!\!
\int dz_1 dz_2 e^{-q|z_1-z_2|} |\zeta(z_1)|^2|\zeta(z_2)|^2,\nonumber\\
&&{v}_{sd}({q})=\frac{e^2}{\kappa l_Bq} \int\! dze^{-q|z-z_d|}|
\zeta(z)|^2\,.
\end{eqnarray}
$\left(\mbox{The impurity cite is assumed to be at the point}\;\; {\bf
R}_d=\{0,0,z_d\}\,.\right)$ \vspace{-2mm}

\section{SPIN OPERATORS}
\label{B}

Bound spin excitons are characterized by the spin numbers $S_z$ and
$S^2$. The corresponding  operators have the form
\begin{equation}\label{II.3}
{\hat S}_z = {\hat S}_z^{(s)} + {\hat S}_z^{(d)}\,,
\end{equation}
where
\begin{equation}\label{II.1}
{\hat S}_z^{(s)}=\frac{N_\phi}{2}\left({\cal A}_0-{\cal
B}_0\right)~; ~~~~\left(\hat{\bf S}^{(s)}\right)^2=N_\phi{\cal
Q}_0^\dag{\cal Q}_0+\left({\hat S}_z^{(s)}\right)^2+{\hat S}_z^{(s)}
\end{equation}
and
\begin{equation}\label{II.2}
{\hat S}_z^{(d)}=\frac{1}{2}({\hat n}_\uparrow-{\hat
n}_\downarrow)~; ~~~~\left(\hat{\bf
S}^{(d)}\right)^2=\frac{3}{4}({\hat n}_\uparrow+{\hat
n}_\downarrow)-\frac{3}{2}{\hat n}_\uparrow{\hat n}_\downarrow
\end{equation}
are the spin operators for 2DEG (in the excitonic representation)
and for magnetic impurity (in terms of the single-orbital model),
respectively. The total squared spin operator for the system is
defined as
\begin{equation}\label{II.4}
\hat{\bf S}^2 = \left(\hat{\bf S}^{(s)}\right)^2 + 2 {\hat
S}_z^{(s)}{\hat S}_z^{(d)} + N_\phi^{1/2}\left({\cal Q}^{\dag}_0
c^\dag_\uparrow c_\downarrow+c^\dag_\downarrow c_\uparrow{\cal
Q}_0\right) + \left(\hat{\bf S}^{(d)}\right)^2~ .
\end{equation}
The operator ${\hat S}_z$ commutes with the Hamiltonian
(\ref{H-tot}), while for $\hat{\bf S}^2$ one has
\begin{equation}\label{II.5}
\begin{array}{l}\left[\hat{\bf S}^2,{\hat {\cal H}}\right] \equiv
N_\phi^{1/2} \left(g_i\mu_BB-\varepsilon_{\rm Z}\right)
c_\downarrow^\dag c_\uparrow{\cal Q}_0\vspace{2mm}
\\
\displaystyle{ + (\beta_\uparrow -
\beta_\downarrow)V\left[c_\downarrow^\dag \left({\hat n}_\uparrow{
b}_0 + N_\phi^{1/2}{\cal Q}_0{ a}_0\right) - c_\uparrow^\dag
\left({\hat n}_\downarrow{ a}_0+N_\phi^{1/2} {\cal Q}_0^\dag {
b}_0\right)\right] - \mbox{H.c.}}
  \end{array}
\end{equation}
The difference between the $g$ factors of the magnetic impurity and
the host QHF and the difference between the projection factors
$\beta_\uparrow$ and $\beta_\downarrow$ measure the spin
non-conservation.


\begin{thebibliography}{99}
\bibitem{foot6} The first term of Eq (\ref{3.10}) is irrelevant
because now the reorientation Zeeman energy is determined not by
reversing the impurity spin but rather by a massive spin-flip of LL
electrons in the vicinity of the impurity. This part of energy is
included in the skyrmion Zeeman energy (\ref{K}).
\bibitem{foot7} Note that due to smallness of the
$F_{\underline{\rm x}}^{(5/2 - \delta S_z)}$ values in the region of
the blank sector parameters in Fig. \ref{f.6}, the $\delta S_z$
peaks for transitions to the continuous spectrum will probably
almost coincide with the $\delta S_z\!+\!1$ peaks related to the
mentioned transitions to the localized states.
\end{thebibliography}
\end{document}